\PassOptionsToPackage{unicode}{hyperref}
\PassOptionsToPackage{hyphens}{url}
\PassOptionsToPackage{dvipsnames,svgnames,x11names}{xcolor}
\documentclass[
  12pt]{article}

\usepackage{amsmath,amssymb}
\usepackage{amsthm}
\usepackage{natbib}
\usepackage{graphicx}
\usepackage{url}
\usepackage{authblk}
\usepackage{enumitem}
\usepackage{multirow}
\usepackage{caption}
\usepackage{subcaption}
\usepackage[section]{placeins} 
\usepackage{booktabs}
\usepackage{algorithm}
\usepackage[noEnd=true]{algpseudocodex}
\usepackage{pdfpages}

\usepackage{iftex}
\ifPDFTeX
  \usepackage[T1]{fontenc}
  \usepackage[utf8]{inputenc}
  \usepackage{textcomp} 
\else 
  \usepackage{unicode-math}
  \defaultfontfeatures{Scale=MatchLowercase}
  \defaultfontfeatures[\rmfamily]{Ligatures=TeX,Scale=1}
\fi
\usepackage{lmodern}
\ifPDFTeX\else  
\fi
\IfFileExists{upquote.sty}{\usepackage{upquote}}{}
\IfFileExists{microtype.sty}{
  \usepackage[]{microtype}
  \UseMicrotypeSet[protrusion]{basicmath} 
}{}
\makeatletter
\@ifundefined{KOMAClassName}{
  \IfFileExists{parskip.sty}{%
    \usepackage{parskip}
  }{
    \setlength{\parindent}{0pt}
    \setlength{\parskip}{6pt plus 2pt minus 1pt}}
}{
  \KOMAoptions{parskip=half}}
\makeatother
\usepackage{xcolor}
\makeatletter
\ifx\paragraph\undefined\else
  \let\oldparagraph\paragraph
  \renewcommand{\paragraph}{
    \@ifstar
      \xxxParagraphStar
      \xxxParagraphNoStar
  }
  \newcommand{\xxxParagraphStar}[1]{\oldparagraph*{#1}\mbox{}}
  \newcommand{\xxxParagraphNoStar}[1]{\oldparagraph{#1}\mbox{}}
\fi
\ifx\subparagraph\undefined\else
  \let\oldsubparagraph\subparagraph
  \renewcommand{\subparagraph}{
    \@ifstar
      \xxxSubParagraphStar
      \xxxSubParagraphNoStar
  }
  \newcommand{\xxxSubParagraphStar}[1]{\oldsubparagraph*{#1}\mbox{}}
  \newcommand{\xxxSubParagraphNoStar}[1]{\oldsubparagraph{#1}\mbox{}}
\fi
\makeatother

\usepackage{longtable,booktabs,array}
\usepackage{calc} 
\usepackage{etoolbox}
\makeatletter
\patchcmd\longtable{\par}{\if@noskipsec\mbox{}\fi\par}{}{}
\makeatother
\IfFileExists{footnotehyper.sty}{\usepackage{footnotehyper}}{\usepackage{footnote}}
\makesavenoteenv{longtable}
\usepackage{graphicx}
\makeatletter
\def\maxwidth{\ifdim\Gin@nat@width>\linewidth\linewidth\else\Gin@nat@width\fi}
\def\maxheight{\ifdim\Gin@nat@height>\textheight\textheight\else\Gin@nat@height\fi}
\makeatother
\setkeys{Gin}{width=\maxwidth,height=\maxheight,keepaspectratio}
\makeatletter
\def\fps@figure{htbp}
\makeatother

\makeatletter
\@ifpackageloaded{caption}{}{\usepackage{caption}}
\AtBeginDocument{%
\ifdefined\contentsname
  \renewcommand*\contentsname{Table of contents}
\else
  \newcommand\contentsname{Table of contents}
\fi
\ifdefined\listfigurename
  \renewcommand*\listfigurename{List of Figures}
\else
  \newcommand\listfigurename{List of Figures}
\fi
\ifdefined\listtablename
  \renewcommand*\listtablename{List of Tables}
\else
  \newcommand\listtablename{List of Tables}
\fi
\ifdefined\figurename
  \renewcommand*\figurename{Figure}
\else
  \newcommand\figurename{Figure}
\fi
\ifdefined\tablename
  \renewcommand*\tablename{Table}
\else
  \newcommand\tablename{Table}
\fi
}
\@ifpackageloaded{float}{}{\usepackage{float}}
\floatstyle{ruled}
\@ifundefined{c@chapter}{\newfloat{codelisting}{h}{lop}}{\newfloat{codelisting}{h}{lop}[chapter]}
\floatname{codelisting}{Listing}

\makeatother
\makeatletter
\@ifpackageloaded{caption}{}{\usepackage{caption}}
\@ifpackageloaded{subcaption}{}{\usepackage{subcaption}}
\makeatother

\ifLuaTeX
  \usepackage{selnolig}  
\fi
\usepackage[]{natbib}
\usepackage{bookmark}

\IfFileExists{xurl.sty}{\usepackage{xurl}}{} 
\hypersetup{
  pdftitle={Title},
  pdfauthor={Author 1; Author 2},
  pdfkeywords={3 to 6 keywords, that do not appear in the title},
  colorlinks=true,
  linkcolor={blue},
  filecolor={Maroon},
  citecolor={Blue},
  urlcolor={Blue},
  pdfcreator={LaTeX via pandoc}}

\graphicspath{ {./figure/} }

\theoremstyle{plain} 
\newtheorem{assum}{Assumption}
\newtheorem{prop}{Proposition}

\newtheorem{corollary}{Corollary}

\theoremstyle{definition} 

\newcommand\se{\text{s.e.}}
\newcommand\N{\mathcal{N}}
\newcommand\E{\mathbb{E}}
\newcommand\X{\mathcal{X}}
\newcommand\di{\mathrm{d}}

\newcommand\cl{\,{:}\,}
\newcommand\R{\mathbb{R}}

\newcommand{\anon}{1}

\begin{document}

\def\spacingset#1{\renewcommand{\baselinestretch}%
{#1}\small\normalsize} \spacingset{1}


\if1\anon
{
  \title{\bf Modular Markov chain Monte Carlo with application to multimodal sampling}
  \author{Joonha Park\hspace{.2cm}\\
    Department of Mathematics, University of Kansas}
  \date{}
  \maketitle
} \fi

\if0\anon
{
  \bigskip
  \bigskip
  \bigskip
  \begin{center}
    {\LARGE\bf Modular Markov chain Monte Carlo with application to multimodal sampling}
\end{center}
  \medskip
} \fi

\bigskip
\begin{abstract}
  We develop a modular approach to Markov chain Monte Carlo (MCMC) sampling for unnormalized target densities.
  In this approach, Markov chains are constructed in parallel, each constrained to a subset of the target space.
  The Monte Carlo estimates from the constrained chains are then combined with appropriate weights, calculated from the transition probabilities between subsets.
  In addition to the computational advantages arising from its parallelized structure, this modular MCMC approach enables variance reduction for Monte Carlo estimation in settings where sampling from low-density regions is required.
  We develop a central limit theorem–type result for the resulting Monte Carlo estimates and propose a method for estimating their standard errors.
  Furthermore, by applying this modular sampling technique to simulated tempering, we propose a method for Monte Carlo estimation of expectations with respect to multimodal target distributions.
  This approach effectively addresses a well-known challenge of tempering-based methods: sampling efficiency can be greatly reduced when separated modes of the target distribution have different scales.
  We demonstrate the efficiency of the proposed methods through numerical examples, including one arising from Bayesian sparse regression with a spike-and-slab prior.
\end{abstract}

\noindent%
\vfill

\spacingset{1.2} 

\section{Introduction}
Markov chain Monte Carlo (MCMC) is widely used for drawing samples from an unnormalized target density.
Applications arise in various fields, including the generation of posterior samples in Bayesian statistics and the computational simulation of microstates in the physical sciences.
However, because of its sequential nature, MCMC often has limited ability to effectively utilize the large number of computing units available in modern computing environments.

We propose a modular approach to MCMC in which multiple Markov chains are constructed, each focusing on a different part of the target space.
This approach enables natural parallelism and facilitates efficient use of computing clusters.
The Monte Carlo samples from the different chains are then combined using weights that reflect the proximities between the regions.

In addition to supporting parallel implementation, the modular approach can also be used when low-variance Monte Carlo estimation over small-probability regions is required.
This application is particularly useful when the target distribution is multimodal---that is, when regions of relatively high density are separated by regions of low density.
Commonly used MCMC algorithms often exhibit efficient sampling within each mode but infrequent transitions between modes, leading to high-variance estimates of the relative probabilities associated with those modes.
Our modular MCMC approach addresses this issue by constructing Markov chains constrained to approximately unimodal regions.
The probabilities associated with the modes are then estimated by solving an associated eigenvalue problem, rather than by evaluating the proportion of draws obtained from each region.
Precise estimation of these probabilities leads to low-variance estimates of expectations with respect to the target distribution.

One of the main contributions of this paper is the development of a \emph{modular simulated tempering} algorithm that combines the idea of tempering with our modular sampling method.
A particular strength of this approach is that it addresses the well-known issue of slow mixing when the modes of the target distribution have different scales.

Tempering techniques have been widely used for handling multimodality.
The essential idea is to introduce an additional temperature variable that modulates the degree of multimodality in the distribution.
Typically, tempering algorithms exploit the fact that transitions between separated modes become more accessible at higher temperatures.
\citet{swendsen1986replica} introduced parallel tempering (PT), later extended by \citet{geyer1991markov}, while \citet{marinari1992simulated} developed simulated tempering (ST).
Other widely used tempering-based MCMC methods include tempered transitions, introduced by \citet{neal1996sampling}, and the equi-energy sampler, introduced by \citet{kou2006equi}.
Tempered sequential Monte Carlo (TSMC) is another tempering-based method, although its core mechanism is sequential importance sampling rather than Markov chain sampling \citep{neal2001annealed}.

All of these tempering-based methods, however, share the same challenge: they cannot effectively draw samples from all modes of the target distribution when those modes have different scales.
This issue arises because modes with different scales can occupy probability masses that vary substantially across temperature levels.
Our modular simulated tempering approach addresses this challenge by estimating the probabilities of the separated modes at all temperature levels through the eigenvector of an associated stochastic matrix.

The idea of using eigendecomposition to determine the relative weights of stratified MCMC samples has also been used in other methods, such as the eigenvector method for umbrella sampling (EMUS) \citep{thiede2016eigenvector, dinner2020stratification}.
However, our modular MCMC method constructs the parallel Markov chains using a mechanism that differs from that of EMUS.
EMUS uses bias potentials to construct overlapping biased distributions.
As a result, the design of the bias potentials is crucial to the efficiency of the resulting Monte Carlo estimation, but achieving this can be challenging in settings such as multimodal distributions.
By contrast, our modular approach constructs Markov chains constrained to non-overlapping subsets of the space.
Consequently, it uses a more straightforward partition of the space, in which each component approximately corresponds to a distinct modal region.

We summarize the main contributions of this work as follows.
\begin{enumerate}
\item We develop a novel, readily parallelizable modular approach to MCMC.
  The parallel Markov chains are constrained to non-overlapping regions of the space, and the samples from these chains are combined using appropriate weights determined by solving an eigenvalue problem.
  Despite some superficial similarities, our modular approach operates through a mechanism that is distinct from that of EMUS.
  In Section~\ref{sec:modmcmc}, we introduce the modular MCMC algorithm and discuss its differences from EMUS and other related methods.
\item We develop theoretical results on the ergodicity and concentration properties of modular MCMC, including a central limit theorem (Section~\ref{sec:clt}).
\item We develop a modular simulated tempering (ST) algorithm for multimodal distributions by combining the modular approach with tempering techniques (Section~\ref{sec:mohmc_st}).
  Notably, this algorithm successfully addresses the torpid mixing often observed in tempering-based methods when separated modes have different scales.
  Moreover, we develop an automatic tuning strategy for modular ST that facilitates efficient Monte Carlo estimation with minimal customization.
\item We develop a method for estimating the standard errors of Monte Carlo estimates obtained using the modular approach (Section~\ref{sec:se_est}).
\item We provide numerical demonstrations of the modular ST algorithm using a toy example with modes of heterogeneous scales and compare its performance with that of adaptive parallel tempering and tempered sequential Monte Carlo algorithms (Section~\ref{sec:numresults}).
  In addition, we demonstrate the practical utility of our method through an applied example of Bayesian sparse regression with a spike-and-slab prior.
\end{enumerate}

\section{Modular Markov chain Monte Carlo}\label{sec:modmcmc}
Consider sampling from a target distribution on a space $\X$ with known unnormalized density $\gamma(x)$.
The corresponding normalized target density is denoted by $\pi(x)$:
\[
  \pi(x) = \frac{\gamma(x)}{Z}.
\]
With a slight abuse of notation, we also use $\pi$ to denote the probability distribution with density $\pi(x)$.
Markov chain Monte Carlo methods generate approximate draws from the target distribution by constructing a Markov chain $\{X^t; t\geq 1\}$ with stationary distribution $\pi$.
Given a Markov chain of length $n$, $(X^1, \dots, X^n)$, the expectation
\[
  \pi h := \E_{X\sim\pi} h(X) = \int_\X h(x) \pi(\di x)
\]
for any $\pi$-integrable function $h$ (that is, $\pi |h| < \infty$) can be estimated by the empirical average,
\[
  \hat \pi h := \frac{1}{n} \sum_{t=1}^n h(X^t).
\]
The variance of the MCMC estimator $\hat \pi h$ depends on the mixing speed of the Markov chain.
For distributions with complex dependence structures or multiple modes, MCMC algorithms often mix slowly, resulting in large Monte Carlo variance.

In this paper, we develop a novel strategy for estimating $\pi h$ with low variance by running multiple Markov chains in parallel.
Consider a partition of the target space $\X$ into $L$ non-overlapping subsets:
\[
  \X = A_1 \sqcup A_2 \sqcup \cdots \sqcup A_L.
\]
Assuming that each subset has positive probability under $\pi$, the constrained distributions on $A_i$ are given by
\[
  \pi_i(\di x) := \frac{\pi(\di x) \mathbf 1_{A_i}(x)}{\pi(A_i)}, \quad i\in 1\cl L.
\]
The original target distribution can then be written as
\[
  \pi = \sum_{i=1}^L \pi(A_i) \pi_i.
\]

For a given $\pi$-integrable function $h$, our modular approach constructs estimates $\hat \pi_i h$ of the conditional expectations $\pi_i h$ and estimates $\widehat{\pi(A_i)}$ of the compartment probabilities $\pi(A_i)$.
The expectation $\pi h$ is then approximated by
\[
  \hat \pi h = \sum_{i=1}^L \widehat{\pi(A_i)} \cdot \hat \pi_i h.
\]
The conditional expectations $\hat \pi_i h$ are estimated by constructing $L$ parallel Markov chains, each constrained to $A_i$ for $i \in \{1,\dots, L\}$.
The compartment probabilities $\pi(A_i)$ are estimated by finding the invariant distribution of a Markov chain defined on the finite space $\{1, \dots, L\}$.
This finite-state Markov chain represents transitions between the compartments $\{A_i\}_{i=1}^L$ under a global Markov kernel $M$ defined on $\mathcal X$.
Rather than explicitly constructing this chain, we estimate its stochastic matrix using the $L$ parallel Markov chains.

\begin{algorithm}[t]
  \caption{Modular Markov chain Monte Carlo}
  \label{alg:mod_mcmc}
  \begin{algorithmic}[1]
    \Statex \textbf{Input:} Partition of the target space, $\X = A_1 \sqcup \cdots \sqcup A_L$; Compartment weights, $\mathbf w = (w_i)_{i\in 1:L}$; Markov transition kernel $M$ that is reversible with respect to $\pi^{\mathbf w}(\di x) = \{\sum_{i=1}^L w_i \mathbf 1[x \in A_i] \pi(\di x)\} / \{\sum_{i=1}^L w_i \pi(A_i)\}$; Function $h$ for which the integral $\pi h$ is to be estimated; Length of constrained Markov chains, $n_i$, $i\in 1\cl L$
    \Statex \textbf{Output:} Monte Carlo estimate $\hat \pi h$ of $\pi h$
    \For {each $i \in 1\cl L$}
      \State Choose $X_i^0$ arbitrarily
      \State Let $C_{ij}^0 \gets 0$ for every $i \neq j$
      \For {$t \in 0\cl (n_i-1)$}
        \State Draw $\tilde X_i^{t+1} \sim M(\cdot; X_i^t)$ \label{line:modmcmc_1}
        \If {$\tilde X_i^{t+1} \in A_i$}
          \State Let $X_i^{t+1} \gets \tilde X_i^{t+1}$
        \EndIf
        \If {$\tilde X_i^{t+1} \in A_j$ with $j\neq i$}
          \State Let $X_i^{t+1} \gets X_i^t$
          \State Let $C_{ij}^{t+1} \gets C_{ij}^t + 1$ \label{line:modmcmc_2}
        \EndIf
      \EndFor
      \State Let $\hat Q_{ij} = C_{ij}^{n_i} / n_i$ for $j \neq i$ and $\hat Q_{ii} = 1 - \sum_{j\neq i} \hat Q_{ij}$
      \State Let $\hat \pi_i h = \tfrac{1}{n_i} \sum_{t=1}^{n_i} h(X_i^t)$
    \EndFor
    \State Let $\hat {\mathbf p}^\mathbf w = (\hat p_1^\mathbf w, \dots, \hat p_L^\mathbf w)_{i\in 1:L}$ be the left eigenvector of $\hat Q$ corresponding to the unit eigenvalue, satisfying $\hat p_j^\mathbf w = \sum_{i=1}^L \hat p_i^\mathbf w \hat Q_{ij}$ for $j \in 1\cl L$
    \State Let $\widehat{\pi(A_i)} = (\hat p_i^\mathbf w/w_i) / (\sum_{j=1}^L \hat p_j^\mathbf w/w_j)$ for $i \in 1\cl L$
    \State Let $\hat \pi h = \sum_{i=1}^L \widehat{\pi(A_i)} \cdot \hat \pi_i h$
  \end{algorithmic}
\end{algorithm}

All $L$ constrained chains, corresponding to $A_i$ for $i\in \{1, \dots, L\}$, are constructed using the same global Markov kernel $M$.
We first define a weight-adjusted target distribution
\[
  \pi^\mathbf{w} := \frac{\sum_{i=1}^L w_i \pi(A_i) \pi_i}{\sum_{i=1}^L w_i \pi(A_i)},
\]
where $\mathbf w = (w_i)_{i\in 1:L}$ is a vector of positive weights.
If $w_i = 1$ for every $i$, $\pi^\mathbf w = \pi$.
We introduce this weight adjustment to improve numerical accuracy when estimating the invariant distribution of the stochastic matrix associated with the finite-state Markov chain on $\{1,\dots, L\}$ using an eigenvector-based method.
The issue of numerical accuracy will be discussed in greater detail when we introduce modular simulated tempering in Section~\ref{sec:mohmc_st}.

The global Markov kernel $M$ is selected to be reversible with respect to $\pi^\mathbf w$:
\[
  M(\di x; y) \pi^\mathbf w(\di y) = M(\di y; x) \pi^\mathbf w(\di x).
\]
The $i$-th Markov chain targeting the conditional distribution $\pi_i$ is constructed as follows.
We denote the state of the Markov chain at the $t$-th iteration by $X_i^t$.
First, a draw is generated from the global Markov kernel conditional on $X_i^t$:
\[
  \tilde X_i^{t+1} \sim M(\cdot; X_i^t)
\]
If the draw $\tilde X_i^{t+1}$ is in $A_i$, then it is taken as the next state of the Markov chain, and we let $X_i^{t+1} = \tilde X_i^{t+1}$.
If $\tilde X^{t+1}$ is in $A_j$ for some $j \neq i$, the Markov chain state does not change: $X_i^{t+1} = X_i^t$.
In this case, we increment the transition counter from $A_i$ to $A_j$ by one: $C_{ij}^{t+1} = C_{ij}^t$.
The $i$-th constrained Markov kernel $M_i$ can be expressed as follows:
\begin{equation}
  M_i(\di x; x_i) = M(\di x; x_i) \mathbf 1[x\in A_i] + M(A_i^{\mathsf c}; x_i) \delta_{x_i} (\di x),
  \label{eqn:Mi}
\end{equation}
where $\delta_{x_i}(\di x)$ denotes the probability measure with all mass concentrated at point $x_i$.
The counters $C_{ij}^t$ for $i \neq j$ are used to estimate the stochastic matrix of the implicit finite-state Markov chain on $\{1,\dots, L\}$.
Algorithm~\ref{alg:mod_mcmc} summarizes the procedure of constructing parallel Markov chains and estimating the compartment probabilities.

Proposition~\ref{prop:constrained} shows that the constrained chains $(X_i^t)_{t\geq 1}$ for $i\in 1\cl L$ have the constrained probability distributions $\pi_i$ as invariant distributions.
Proofs of all propositions are provided in the supplementary material.
\begin{prop}\label{prop:constrained}
  For each $i \in 1\cl L$, the constrained chain $M_i$ constructs a reversible chain supported on $A_i$ having $\pi_i$ as an invariant distribution.
\end{prop}

Let $c_j(x_i) = M(A_j; x_i)$ be the transition probability from $x_i \in A_i$ to $A_j$.
Define a matrix $Q \in \R^{L\times L}$ such that
\begin{equation}
  Q_{ij} = \pi_i(c_j) = \int M(A_j; x_i) \pi_i(\di x_i) = P_\pi[\tilde X^{t+1}_i \in A_j | X^t_i \in A_i]
  \label{eqn:Q}
\end{equation}
for $i\neq j$ and $Q_{ii} = 1-\sum_{j\neq i} Q_{ij}$.
Note that both $\pi$ and $\pi^\mathbf w$ yield the same conditional distribution $\pi_i$ when constrained to $A_i$.
Thus for $i \neq j$, $Q_{ij}$ gives the average transition probability from $A_i$ to $A_j$ under the Markov kernel $M$, provided that $X_i^t$ is in $A_i$ under the stationary distribution $\pi^\mathbf w$.
The stochastic matrix $Q$ implicitly depends on the weights $\mathbf w$ through $M$.

For each $i \in \{1,\dots, L\}$, let
\[
  p_i^\mathbf w := \pi^\mathbf w(A_i) = \frac{w_i \pi(A_i)}{\sum_{j=1}^L w_j \pi(A_j)}.
\]
Proposition~\ref{prop:Qinv} shows that $\mathbf p^\mathbf w = (p_i^\mathbf w)_{i\in 1\cl L}$ is an invariant distribution of the finite-state Markov chain with stochastic matrix $Q$.
\begin{prop}\label{prop:Qinv}
  $\mathbf p^\mathbf w = (p_i^\mathbf w)_{i\in 1:L}$ is the left eigenvector of $Q$ corresponding to the unit eigenvalue, that is, $(\mathbf p^\mathbf w)^\top Q = (\mathbf p^\mathbf w)^\top$.
\end{prop}

Suppose that a Markov chain of length $n_i$ has been constructed constrained on $A_i$.
Then we estimate the $i$-th row of the stochastic matrix $Q$ through 
\[
  \hat Q_{ij} := \frac{C_{ij}^{n_i}}{n_i}, \quad i\neq j
\]
and $\hat Q_{ii} = 1 - \sum_{j\neq i} \hat Q_{ij}$.
$\hat Q_{ij}$ is an unbiased estimate of $Q_{ij} = \pi_i(c_j)$, provided that the Markov chain constrained to $A_i$ starts from the stationary distribution $\pi_i$.

\begin{algorithm}[t]
  \caption{Modular MCMC using a Metropolis-Hastings type global kernel}
  \label{alg:MHcounter}
  \begin{algorithmic}[1]
    \Statex \textbf{Input:} Same as the input for Algorithm~\ref{alg:mod_mcmc}, but the global Markov kernel $M$ is constructed using a proposal kernel with density $m(\cdot;\cdot)$ (Equation~\ref{eqn:MH_base})
    \Statex \textit{Replace lines \ref{line:modmcmc_1}--\ref{line:modmcmc_2} of Algorithm~\ref{alg:mod_mcmc} by the following}
    \State Draw $X_i^\text{cand} \sim m(\cdot; X_i^t)$ 
    \If {$X_i^\text{cand} \in A_i$}
      \State Draw $U \sim \text{Unif}[0,1]$
      \If {$U < \frac{\pi(X_i^\text{cand}) m(X_i^t; X_i^\text{cand})}{\pi(X_i^t) m(X_i^\text{cand}; X_i^t)}$}
        \State Let $X_i^{t+1} \gets X_i^\text{cand}$
      \Else
        \State Let $X_i^{t+1} \gets X_i^t$
      \EndIf
    \EndIf
    \If {$X_i^\text{cand} \in A_j$ with $j\neq i$}
      \State Let $X_i^{t+1} \gets X_i^t$
      \State Let $C_{ij}^{t+1} \gets C_{ij}^t + \min\left(1, \frac{\pi(X_i^\text{cand}) w_j \cdot m(X_i^t; X_i^\text{cand})}{\pi(X_i^t) w_i \cdot m(X_i^\text{cand}; X_i^t)} \right)$
    \EndIf
  \end{algorithmic}
\end{algorithm}

If the global Markov kernel $M$ is constructed using the Metropolis-Hastings (MH) strategy, the variance of the estimate $\hat Q_{ij}$ can be reduced by utilizing a conditional expectation.
Suppose that we draw a candidate state $X^\text{cand}$ from a proposal kernel with density $m(\cdot; X)$, where $X$ denotes the current state of the Markov chain.
The candidate is then accepted with probability
\[
  \alpha(X^\text{cand}; X) = \min\left(1, \frac{\pi^{\mathbf w}(X^\text{cand})}{\pi^{\mathbf w}(X)} \frac{m(X; X^\text{cand})}{m(X^\text{cand}; X)} \right),
\]
and if it is rejected, the Markov chain remains at state $X$.
The resulting Markov kernel
\begin{equation}
  M(\di x'; x) = \alpha(x'; x) m(x'; x) \di x' + \left( 1- \int_{\mathcal X} \alpha(x'; x) m(x'; x) \di x' \right) \delta_x (\di x')
  \label{eqn:MH_base}
\end{equation}
is reversible with respect to the unnormalized density $\pi(x)$ \citep{hastings1970monte}.
If a Metropolis-Hastings kernel is used in our modular approach as a global kernel, the state of the Markov chain constrained on $A_i$ changes only when $X^\text{cand}$ is in $A_i$ and accepted.
If $X^\text{cand}$ is in a different compartment $A_j$, we do not decide whether it is accepted, but rather increase the counter $C_{ij}$ by the acceptance probability $\alpha(X^\text{cand}; X)$.
The expected increase in $C_{ij}$ conditional on the current state $X$ is given by
\[
  \int \alpha(x^\text{cand}; X) \mathbf 1[x^\text{cand} \in A_i] m(x^\text{cand}; X) \di x^\text{cand},
\]
which is the same as that when the original, integer-valued counter is used.
Provided that the chain starts from the stationary distribution $\pi_i$, the estimator $\hat Q_{ij} = C_{ij}^{n_i} / n_i$ using the modified counter is still unbiased for $Q_{ij}$ and has a variance that is less than or equal to that of the original estimator due to Blackwell's theorem \citep{blackwell1947conditional}.
Thus, using this modified counter is recommended whenever the global kernel is of an MH type.
This procedure is summarized in Algorithm~\ref{alg:MHcounter}.

We discuss the case where the global Markov kernel $M$ is constructed using Hamiltonian Monte Carlo (HMC).
HMC is a widely used MCMC algorithm due to its favorable mixing and dimension scaling properties \citep{duane1987hybrid, neal2011mcmc}.
It proposes a candidate for the next state of the Markov chain by numerically simulating the Hamiltonian dynamics.
The momentum of the simulated particle changes at a rate determined by the gradient of the logarithm of the target density.
The initial momentum is drawn from a distribution with density $\psi$, which is often taken to be the multivariate normal distribution.
The trajectory is numerically constructed using the leapfrog method, and the end point is either accepted or rejected according to a Metropolis-Hastings ratio.
When using HMC within modular MCMC, the Hamiltonian trajectories can be constructed using the gradient of the logarithm of the weight-unadjusted target density $\pi$ to avoid discontinuity when moving from one compartment to another.
However, the acceptance probability should take into account the compartment weights: if the current state of the Markov chain is $x \in A_i$ and the end point is $x^\text{cand} \in A_j$ and the initial and the final momenta are denoted by $\rho$ and $\rho^\text{cand}$ respectively, the acceptance probability is given by
\[
  \alpha(x^\text{cand}, \rho^\text{cand}; x, \rho) = \min\left(1, \frac{w_j \pi(x^\text{cand}) \psi(\rho^\text{cand})}{w_i \pi(x) \psi(\rho)} \right) = \min\left(1, \frac{\pi^{\mathbf w}(x^\text{cand}) \psi(\rho^\text{cand})}{\pi^{\mathbf w}(x) \psi(\rho)} \right).
\]
Nonetheless, the resulting constrained Markov chain on each $A_i$ is reversible with respect to $\pi_i$ because the numerical simulation of Hamiltonian trajectories is itself reversible (see, e.g., \citet{park2020markov}).

Let $\hat{\mathbf p}^\mathbf w = (\hat p_i^\mathbf w)_{i \in 1\cl L}$ be a left eigenvector of $\hat Q$ corresponding to the unit eigenvalue:
\[
  (\hat{\mathbf p}^\mathbf w)^\top \hat Q = (\hat{\mathbf p}^\mathbf w)^\top.
\]
There are several methods for computing the eigenvector of a stochastic matrix corresponding to the unit eigenvalue \citep{hunter1982generalized, hunter1991computation}.
In all of our numerical examples, we computed the eigenvector of $\hat Q$ corresponding to the unit eigenvalue using the method described in \citet{golub1986using}: perform the QR decomposition of $\hat A := I - \hat Q$ and take the right-most column of the unitary matrix in the factorization.
The compartment probabilities $\pi(A_i)$ for $i \in \{1,\dots, L\}$ are then estimated by dividing the entries of $\mathbf p^\mathbf w$ by the corresponding weights:
\[
  \widehat{\pi(A_i)} = \frac{\hat p_i^\mathbf w/w_i}{\sum_{j=1}^L \hat p_j^\mathbf w / w_j}.
\]

If $\{X_i^t; 1\leq t \leq n_i\}$ denotes the Markov chain on $A_i$, then for any $\pi$-integrable function $h$, we estimate the constrained expectation of $h$ through
\[
  \hat \pi_i h = \frac{1}{n_i} \sum_{t=1}^{n_i} h(X_i^t).
\]
The expectation of $h$ with respect to the target distribution $\pi$ is then estimated by
\begin{equation}
  \hat \pi h = \sum_{i=1}^L \widehat{\pi(A_i)} \cdot \hat \pi_i h.
  \label{eqn:pihat}
\end{equation}
Algorithm~\ref{alg:mod_mcmc} summarizes the procedure for estimating $\pi h$ using modular MCMC.

\subsection{Comparison with other stratified sampling methods}\label{sec:related_methods}

Modular MCMC (Algorithms~\ref{alg:mod_mcmc} and \ref{alg:MHcounter}) reduces the variance of Monte Carlo estimates through stratification.
\citet{thiede2016eigenvector, dinner2020stratification} proposed another stratified sampling method, called EMUS, and considered applications in Bayesian inference and in free energy computation in computational physics.
Their work addressed sampling from multimodal distributions and estimation of tail probabilities.
EMUS is similar to the modular approach presented in this paper in that both solve an eigenvalue problem to estimate the relative weights of stratified samples.
EMUS computes these weights by constructing unnormalized densities that have overlaps with each other to estimate the ratios of the normalizing constants.
Umbrella sampling \citep{torrie1977nonphysical, chen1997monte},  bridge sampling \citep{bennet1976efficient, meng1996simulating}, and path sampling \citep{gelman1998simulating} are methods for estimating the ratio of normalizing constants for two distributions using Monte Carlo draws, but EMUS extends this idea to estimate the ratios of normalizing constants of more than two distributions.

EMUS samples from biased distributions
\begin{equation}
  \pi_i(\di x) = \frac{\psi_i(x) \pi(\di x)}{\pi(\psi_i)}
  \label{eqn:emus}
\end{equation}
where $\pi$ is the target distribution and $\{\psi_i\}_{i=1}^L$ are nonnegative bias functions, satisfying $\sum_{i=1}^L \psi_i(x) > 0$ for every $x \in \mathcal X$.
For a function $h$ whose expectation with respect to $\pi$ is to be computed, an estimate of $\pi_i(h/\sum_{j=1}^L \psi_j)$ is computed using MCMC draws targeting $\pi_i$ for each $i\in 1\cl L$.
The expectation of $h$ with respect to $\pi$ is then estimated based on the identity
\[
  \pi(h) = \frac{\sum_{i=1}^L \pi(\psi_i) \pi_i \left( \frac{h}{\sum_{j=1}^L \psi_j} \right)}{\sum_{i=1}^L \pi(\psi_i) \pi_i\left( \frac{1}{\sum_{j=1}^L \psi_j}\right)}.
\]
The relative weights $\pi(\psi_i)$ are computed as the left eigenvector corresponding to the unit eigenvalue of a stochastic matrix $F$ defined by
\[
  F_{ij} = \pi_i\left( \frac{\psi_j}{\sum_{k=1}^L \psi_k} \right).
\]
To ensure that $F$ is irreducible and that the Monte Carlo estimate of $\pi(h)$ has low variance, the biased distributions should overlap sufficiently.
In addition, the bias functions must be chosen such that MCMC sampling from each biased distribution is efficient.
Designing such bias functions in practice can be difficult.

By contrast, modular MCMC (Algorithms~\ref{alg:mod_mcmc} and \ref{alg:MHcounter}) uses non-overlapping partition $\{A_i\}_{i=1}^L$ of the target space $\mathcal X$.
The expectation $\pi(h)$ is estimated based on the identity $\pi(h) = \sum_{i=1}^L \pi(A_i) \pi_i(h)$, where the constrained distributions
\[
  \pi_i(\di x) = \frac{\mathbf 1_{A_i}(x) \pi(\di x)}{\pi(A_i)}
\]
correspond to the choice of $\psi_i = \mathbf 1_{A_i}$ in \eqref{eqn:emus}.
The probabilities $\pi(A_i)$ are computed as the left eigenvector corresponding to the unit eigenvalue of a different stochastic matrix, $Q$, whose entries are defined by $Q_{ij} = \pi_i(M(A_j, \cdot))$ where $M$ is the global Markov kernel.

For multimodal distributions, the sets $A_i$ may be chosen to correspond to individual modal regions so that each constrained distribution $\pi_i$ is approximately unimodal and therefore readily admits efficient MCMC sampling.
Moreover, in Section~\ref{sec:mohmc_st}, we introduce a global Markov kernel $M$ inspired by simulated tempering, so that the resulting algorithm facilitates frequent transitions between the non-overlapping subsets in an augmented space (see Algorithm~\ref{alg:modular_st}).
Consequently, this algorithm produces low-bias Monte Carlo estimates.

There are other parallel MCMC algorithms that use partitions of the state space.
\citet{basse2016parallel} and \citet{vanderwerken2013parallel} consider estimating the weight of each compartment using methods such as bridge sampling relative to a normalized distribution that approximates the constrained distribution.
The approximating distribution is typically chosen to be one whose density can be readily evaluated, such as a normal distribution, a $t$-distribution, or a mixture of such distributions.
In practice, however, finding good approximating distributions can be challenging when the constrained distributions are highly skewed or irregularly shaped.
Moreover, the approximation may also lead to highly variable weight estimates in high dimensional settings.

\section{Ergodicity and central limit theorem for modular MCMC}\label{sec:clt}

In this section, we develop ergodicity and a central limit theorem for modular Markov chain Monte Carlo.
We make the following assumptions.

\begin{assum}\label{assum:Qirreducible}
  The stochastic matrix $Q$ defined in \eqref{eqn:Q} is irreducible. 
\end{assum}
    
\begin{assum}\label{assum:ir_ap}
  The Markov chain $\{X_i^t; t\geq 1\}$ constrained to each compartment $A_i$ for $i \in 1\cl L$ is $\pi_i$-irreducible, aperiodic, and Harris recurrent.
\end{assum}

\begin{assum}[Geometric drift condition]\label{assum:drift}
  For each $i\in 1\cl L$, the Markov transition kernel $M_i$ constrained to $A_i$ (Equation~\ref{eqn:Mi}) satisfies the following drift condition:
  there exists an extended real-valued function $V_i:A_i \to [1,\infty]$, a petite set $S_i \subset A_i$, $b_i<\infty$, and $\beta_i>0$ such that
  \[
    M_i V_i (x) - V_i(x) \leq -\beta_i V_i(x) + b_i \mathbf{1}[x \in S_i], \quad x \in A_i.
  \]
\end{assum} 

First, we establish the central limit theorem for the transition counters $C_{ij}^t$, which are used to estimate the stochastic matrix $Q$.
The result is developed for both Algorithm~\ref{alg:mod_mcmc} and Algorithm~\ref{alg:MHcounter}.
Proofs of all propositions are provided in the supplementary material.
We note that, when Algorithm~\ref{alg:MHcounter} is used, the global Markov kernel $M$ is given by Equation~\ref{eqn:MH_base}.

\begin{prop}\label{prop:CLTtrans}
  Suppose that Assumptions~\ref{assum:ir_ap} and \ref{assum:drift} hold.
  For each $i \in 1\cl L$, let $(C^t_{ij})_{j\neq i}$ denote the $L-1$ dimensional random vector consisting of the transition counters from $A_i$ to $A_j$ constructed using either Algorithm~\ref{alg:mod_mcmc} or Algorithm~\ref{alg:MHcounter}.
  Let $c_j(x) = M(A_j; x)$ for every $x \in \mathcal X$ where $M$ is the global Markov kernel, and let $Q_{ij} = \pi_i(c_j)$.
  Then the following joint central limit theorem holds as $t\to\infty$:
  \[
    \frac{1}{\sqrt t}\left\{ (C^t_{ij})_{j\neq i} - t \cdot (Q_{ij})_{j\neq i} \right\} \underset{t\to\infty}\Longrightarrow \mathcal N(0, \Sigma_i),
  \]
  where the elements of the variance $\Sigma_i \in \mathbb R^{L-1, L-1}$ are given by
  \begin{equation}
    (\Sigma_i)_{jj'} = \delta_j^{j'} Q_{ij} - Q_{ij} Q_{ij'} + \sum_{l=1}^\infty \left\{\pi_i(c_j \cdot M^{l-1}_i c_{j'} + c_{j'} \cdot M^{l-1}_i c_j) - 2 Q_{ij} Q_{ij'} \right\}
    \label{eqn:Sigma_alg1}
  \end{equation}
  when Algorithm~\ref{alg:mod_mcmc} is used.
  Here $\delta_j^{j'}$ is equal to 1 if $j=j'$ and 0 otherwise.
  When Algorithm~\ref{alg:MHcounter} is used, the first term on the right hand side of \eqref{eqn:Sigma_alg1} is replaced by
  \[
    \delta_j^{j'} \pi_i\left(\int_{A_j} \alpha(x;\cdot)^2 m(x;\cdot) \di x \right)
  \]
  where $\alpha(x;x') = \min(1, \{\pi^\mathbf w(x) m(x';x)\}/\{\pi^\mathbf w(x') m(x; x')\})$.
\end{prop}

As a corollary to Proposition~\ref{prop:CLTtrans}, we obtain the following consistency result.
\begin{corollary}\label{cor:comp_prob_estimates_consistency}
  Suppose that Assumptions~\ref{assum:ir_ap} and \ref{assum:drift} hold. Then for each $i \in 1\cl L$, $(\hat Q_{ij})_{j\in 1:L}$ converge in probability to $(Q_{ij})_{j\in 1:L}$ as the length of the constructed Markov chain $n_i$ tends to infinity.
\end{corollary}

We will consider a scenario where the lengths of the constrained chains $n_i$ grow linearly with a common variable $\tau$, which represents the time it took to construct the $L$ parallel chains.
Specifically, we assume that the lengths $n_i(\tau)$ depend on $\tau$ such that
\[
  \lim_{\tau\to\infty} \frac{n_i(\tau)}{\tau} = \nu_i \in (0,\infty)
\]
for each $i$.
Under this assumption, Proposition~\ref{prop:CLTest} establishes the asymptotic normality of the estimate $\hat \pi h$ constructed using modular MCMC.

\begin{prop}\label{prop:CLTest}
  Suppose that Assumptions~\ref{assum:Qirreducible}--\ref{assum:drift} hold and that $h:\X \to \R$ is a function whose restriction to each compartment satisfies $(h|_{A_i})^2 \leq V_i$ for each $i\in 1\cl L$ with $V_i$ defined in Assumption~\ref{assum:drift}.
  Assume that the lengths of the constrained chains $n_i$ increase linearly with a length variable $\tau$ for all $i$.
  Then the Monte Carlo estimate $\hat \pi h$ given by \eqref{eqn:pihat} satisfies the central limit theorem:
  \[
    \sqrt \tau (\hat \pi h - \pi h) \underset{\tau\to\infty}\Longrightarrow \mathcal N(0, \sigma_h^2)
  \]
  for some $0 < \sigma_h^2 < \infty$.
\end{prop}

The following corollary summarizes the consistency results for the Monte Carlo estimate $\hat \pi h$ as well as those for the intermediate estimates $\widehat{\pi(A_i)}$ and $\hat \pi_i h$ for $i\in 1\cl L$.

\begin{corollary}\label{cor:MCest_consitency}
  Under the conditions of Proposition~\ref{prop:CLTest}, $(\widehat{\pi(A_i)})_{i\in 1:L}$, $(\hat \pi_i h)_{i\in 1:L}$, and $\hat \pi h$ converge in probability to $(\pi(A_i))_{i\in 1:L}$, $(\pi_i h)_{i\in 1:L}$, and $\pi h$, respectively.
\end{corollary}

\section{Modular simulated tempering for multimodal sampling}\label{sec:mohmc_st}

\begin{algorithm}[t]
  \caption{Modular simulated tempering}
  \label{alg:modular_st}
  \begin{algorithmic}[1]
    \Statex \textbf{Input:} Partition of the target space, $\X = A_1 \sqcup \cdots \sqcup A_L$; Inverse temperature levels, $\{\beta_k: k\in 0\cl K\}$; Mixture weights, $\{w_{k,i}: k\in 0\cl K, i\in 1\cl L\}$; Density of a base distribution, $q$; Markov kernel for state moves $M^\text{state}(\di x'; k,x)$ that is reversible with respect to the conditional density $\pi^\text{aug}(x|k)$ for each fixed $k \in 0\cl K$ (see Equation~\ref{eqn:piaug}); Function $h$ for which the integral $\pi h$ is to be estimated; Length of constrained Markov chains, $n_{k,i}$ for $k \in 0\cl K$ and $i\in 1\cl L$
    \Statex \textbf{Output:} Monte Carlo estimate $\hat \pi h$ of $\pi h$
    \For {each $k \in 0\cl K$ and $i \in 1\cl L$}
      \State Choose $X_{k,i}^0$ arbitrarily
      \State Let $C_{(k,i) \to (k',j)}^0 \gets 0$ for every $k' \in 0\cl K$ and $j \in 1\cl L$
      \For {$t \in 0\cl (n_{k,i}-1)$}
        \State Draw $\tau \sim \text{Bernoulli}(\tfrac{1}{2})$ \Comment{to determine state or temperature move}
        \If {$\tau = 0$} \Comment{state move}
          \State Draw $\tilde X_{k,i}^{t+1} \sim M^\text{state}(\cdot; k, X_{k,i}^t)$
          \If {$\tilde X_{k,i}^{t+1} \in A_i$}
            \State Let $X_{k,i}^{t+1} \gets \tilde X_{k,i}^{t+1}$
          \EndIf 
          \If {$\tilde X_{k,i}^{t+1} \in A_j$ with $j\neq i$}
            \State Let $X_{k,i}^{t+1} \gets X_{k,i}^t$
            \State Let $C_{(k,i)\to(k,j)}^{t+1} \gets C_{(k,i)\to(k,j)}^t + 1$
           \EndIf
         \EndIf
        \If {$\tau = 1$} \Comment{temperature move}
          \State Draw $\Delta k \in \{1, -1\}$ with probability $\tfrac{1}{2}$ each
          \State Let $k' = \min( \max(k + \Delta k, 0), K)$
          \State Let $X_{k,i}^{t+1} \gets X_{k,i}^t$
          \State Let $C_{(k,i) \to (k',i)}^{t+1} \gets C_{(k,i)\to (k',i)}^t + \alpha(k'; k, x)$ where $\alpha(k'; k,x)$ is given by Equation~\ref{eqn:accprob_temp}
        \EndIf
      \EndFor
      \State Let $\hat Q_{(k,i)\to(k',j)} = C_{(k,i)\to(k',j)}^{n_{k,i}} / n_{k,i}$ for $(k',j) \neq (k,i)$ and $\hat Q_{(k,i)\to(k,i)} = 1 - \sum_{(k',j)\neq (k,i)} \hat Q_{(k,i)\to(k',j)}$
      \State Let $\hat \pi_{k,i} h = \tfrac{1}{n_{k,i}} \sum_{t=1}^{n_{k,i}} h(X_{k,i}^t)$
    \EndFor
    \State Let $\hat {\mathbf p}^\mathbf w = (\hat p_{k,i}^\mathbf w)_{k \in 0\cl K, i\in 1\cl L}$ be the left eigenvector of $\hat Q$ corresponding to the unit eigenvalue, satisfying $\hat p_{k',j}^\mathbf w = \sum_{k=0}^K \sum_{i=1}^L \hat p_{k,i}^\mathbf w \hat Q_{(k,i)\to(k',j)}$ for $k' \in 0\cl K$ and $j \in 1\cl L$
    \State Let $\widehat{\pi(A_i)} = (\hat p_{K,i}^\mathbf w/w_{K,i}) / (\sum_{j=1}^L \hat p_{K,j}^\mathbf w/w_{K,j})$ for $i \in 1\cl L$
    \State Let $\hat \pi h = \sum_{i=1}^L \widehat{\pi(A_i)} \cdot \hat \pi_{K,i} h$
  \end{algorithmic}
\end{algorithm}

We develop a method for generating a weighted sample from multimodal target distributions using the modular approach developed in Section~\ref{sec:modmcmc}.
This is achieved by augmenting the state space with an additional temperature dimension, as in simulated tempering \citep{marinari1992simulated, geyer1995annealing}.
Simulated tempering targets a mixture of tempered distributions
\[
  \sum_{k=0}^K w_k \left(\frac{\pi(x)}{q(x)}\right)^{\beta_k} q(x),
\]
where $\beta_k$ denotes the $k$-th inverse temperature level and $q(x)$ the density of a base distribution.
The zeroth inverse temperature level, $\beta_0 = 0$, corresponds to the base distribution, while the highest level, $\beta_K = 1$, the original target density $\pi$.
However, if the mixture weights $w_k$ are not properly tuned, moves between temperature levels may occur very infrequently, resulting in a disproportionately small number of draws from some modes.
To address this issue, adaptive methods for tuning the mixture weights have been developed \citep{wang2001efficient, atchade2010wang}.

Another challenge faced by tempering-based methods for multimodal sampling is slow mixing when separated modes have different scales.
This arises because the modes can occupy substantially different probability masses across temperature levels \citep{woodard2009sufficient, woodard2009conditions}.
Widely used tempering methods, including simulated tempering, parallel tempering, and tempered sequential Monte Carlo, all share this same challenge.
This issue has limited the practical applicability of tempering methods to a broad range of problems.

Our modular MCMC approach, combined with tempering, can address these issues.
First, unlike existing tempering methods, in which few or no sample points may be drawn from some modes at relatively high temperature levels, our modular method ensures that a sufficient number of samples are drawn from each mode and every temperature level.
As a result, the resulting Monte Carlo estimates are less sensitive to the tuning of the mixture weights, since the compartment probabilities are computed algebraically by finding the eigenvector of the estimated stochastic matrix.
Second, our modular approach can assign different mixture weights to different modes at each temperature level.
This feature enables reliable Monte Carlo estimation of expectations even when the modes have different scales.

Our modular approach extends the target space to $\{0, \dots, K\} \times \mathcal X$, similarly to simulated tempering, where the index $k \in \{0,\dots, K\}$ specifies the inverse temperature level.
Let $A_1, \dots, A_L$ be the compartments partitioning the space $\mathcal X$.
Then we consider the mixture distribution with density
\begin{equation}
  \pi^\text{aug}(k, x) \propto \sum_{i=1}^L w_{k,i} \left(\frac{\gamma(x)}{q(x)}\right)^{\beta_k} q(x) \mathbf 1[x \in A_i].
  \label{eqn:piaug}
  \end{equation}
The base distribution with density $q(x)$ is chosen to be readily sampled from and to have a sufficient width to include all possible mode locations.
The modular simulated tempering method constructs a constrained Markov chain for each pair $(k,i) \in \{0,\dots, K\} \times \{1, \dots, L\}$.
In Section~\ref{sec:tuning}, we propose a method to tune $\{\beta_k\}$ and the mixture weights $\{w_{k,i}\}$ to enhance the numerical accuracy in finding the eigenvector of the associated stochastic matrix---numerical inaccuracy can arise when some matrix elements are very small due to finite machine precision.

We use an augmented Markov transition kernel defined on the extended space $\{0, \dots, K\} \times \mathcal X$ consisting of the state transition kernel and the temperature transition kernel:
\[
  M^\text{aug}(k', \di x'; k, x) = \frac{1}{2} M^\text{state}(\di x'; k,x) \mathbf 1[k' = k] + \frac{1}{2} M^\text{temp}(k'; k, x) \delta_x(\di x').
\]
The state transition kernel $M^\text{state}$ can be any Markov kernel, including the HMC kernel discussed in Section~\ref{sec:modmcmc}, that is reversible with respect to the conditional density of $\pi^\text{aug}(k,x)$ for given $k$.
For the temperature transition kernel $M^\text{temp}$, we use a Metropolis-Hastings kernel that proposes temperature moves of one level up or down with equal probabilities:
\begin{multline*}
  M^\text{temp}(k'; k, x) = \frac{1}{2} \alpha(k'; k, x) \mathbf 1[k' = \max(k-1, 0)]
  + \frac{1}{2} \alpha(k'; k, x) \mathbf 1[k' = \min(k+1, K)] \\
  + \frac{1}{2} \{ 2-\alpha(\max(k-1, 0); k,x)-\alpha(\min(k+1,K);k,x) \} \mathbf 1[k' = k],
\end{multline*}
where the acceptance probability for the proposed move is given by
\begin{equation}
  \alpha(k'; k, x) = \min\left\{ 1, \frac{w_{k',i}}{w_{k,i}} \left(\frac{\gamma(x)}{q(x)}\right)^{\beta_{k'} - \beta_k} \right\}
  \label{eqn:accprob_temp}
\end{equation}
provided that $x \in A_i$.
Since we are constructing a chain constrained on $\{k\} \times A_i$, there is no change in the Markov chain state if the temperature move is selected.
However, the acceptance probability \eqref{eqn:accprob_temp} is used to increase the appropriate transition counter.

Modular simulated tempering arises naturally by applying modular MCMC to the extended space $\{0,\dots, K\} \times \mathcal X$ using the augmented Markov kernel $M^\text{aug}$ as the global kernel.
We construct a Markov chain constrained in the set $\{k\} \times A_i$ for each $i \in \{1, \dots, L\}$ and $k \in \{0, \dots, K\}$.
In each constrained Markov chain, the state $(k, x)$ remains unchanged when the draw from $M^\text{aug}(k', \di x'; k, x)$ is outside its support, $\{k\} \times A_i$.
The transition counters are updated as follows.
If the state kernel is selected at the current iteration (with probability $\tfrac{1}{2}$) and the draw is in $\{k\} \times A_j$ with $j \neq i$, the counter $C_{(k,i)\to(k,j)}$ increases by one.
However, if $M^\text{state}$ is an MH kernel and the proposal kernel has density $m(x'; x)$, then we may instead increase the counter as
follows: provided that the current state of the chain and the proposed candidate are denoted by $X_{k,i}^t$ and $X_{k,i}^\text{cand} \in A_j$, respectively,
\[
  C_{(k,i)\to(k,j)} \gets C_{(k,i)\to(k,j)} + \min\left(1, \frac{\gamma(X_{k,i}^\text{cand}) w_{k,j} \cdot m(X_{k,i}^t;X_{k,i}^\text{cand})}{\gamma(X_{k,i}^t) w_{k,i} \cdot m(X_{k,i}^\text{cand};X_{k,i}^t)}\right).
\]
In this case, we do not determine whether $X_{k,i}^\text{cand}$ is accepted.
If the temperature kernel is selected, then $k'$ is proposed between $\max(k-1, 0)$ and $\min(k+1, K)$ with equal probability.
If $k' \neq k$, then we increase the counter as
\[
  C_{(k,i)\to(k',i)} \gets C_{(k,i)\to(k',i)} + \alpha(k'; k,x)
\]
where $\alpha(k'; k, x)$ is given by Equation~\ref{eqn:accprob_temp}.

We define $\hat Q$ to be a $(K+1)L \times (K+1)L$ matrix with entries $\hat Q_{(k,i)\to(k',j)} = C_{(k,i)\to(k',j)} / n_{k,i}$ for $(k',j) \neq (k,i)$, where $n_{k,i}$ is the length of the constructed Markov chain constrained to $\{k\} \times A_i$.
The diagonal entries $\hat Q_{(k,i)\to(k,i)}$ are determined such that the all row sums of $\hat Q$ are equal to one.
Denoting by $\hat{\mathbf p} = (\hat p_{k,i})_{k \in 0:K, i\in 1:L}$ the left eigenvector of $\hat Q$ corresponding to the unit eigenvalue, the compartment probabilities $\pi(A_i)$ are estimated by
\[
  \widehat{\pi(A_i)} = \frac{\hat p_{K,i}/w_{K,i}}{\sum_{j=1}^L p_{K,j}/w_{K,j}},
\]
using only the entries of the estimated eigenvector $\hat{\mathbf p}$ corresponding to the $K$-th temperature level.
Finally, the expectation $\pi h$ with respect to the original target distribution is estimated by
\[
  \hat \pi h = \sum_{i=1}^L \widehat{\pi(A_i)} \frac{1}{n_{K,i}} \sum_{t=1}^{n_{K,i}} h(X^t_{K,i}).
\]
Here $X^t_{K,i}$ denotes the $t$-th state of the constrained Markov chain in $\{K\} \times A_i$.
Algorithm~\ref{alg:modular_st} presents pseudocode for modular ST.

In order to reliably estimate the compartment probabilities $\pi(A_i)$ for $i\in \{1,\dots, L\}$, the augmented Markov kernel $M^\text{aug}$ defined on the space $\{0, \dots, K\} \times \mathcal X$ should be numerically irreducible; that is, there should be only a single eigenvalue of the estimated matrix $\hat Q$ that is numerically indistinguishable from the unity.
Modular ST (Algorithm~\ref{alg:modular_st}) facilitates the numerical irreducibility by enabling transitions between any pair of compartments $(A_i, A_j)$ at the top $\beta$ level (i.e., $\beta_K = 1$) through the path
\[
  (K, A_i) ~\to~ (K-1, A_i) ~\to~ \cdots ~\to~ (0, A_i) ~\to~ (0, A_j) ~\to~ (1, A_j) ~\to~ \cdots ~\to~ (K, A_j).
\]
The transition from $(0, A_i)$ to $(0, A_j)$ is ensured to occur with reasonable frequencies by selecting the base distribution $q(x)$ to have a broad support encompassing all likely mode locations.
Additionally, both up and down moves across $\beta$ levels should occur with reasonably high rates for each compartment.
We propose a strategy for tuning $\{\beta_k\}_{k \in 0:K}$ and $\{w_{k,i}\}_{k\in 0:K, i\in 1:L}$ that ensures this in Section~\ref{sec:tuning}.

When an HMC kernel is used for state moves, the step sizes for the numerical simulation of Hamiltonian trajectories need to be tuned appropriately to achieve both numerical stability and computational efficiency.
If the target density is Gaussian, the numerical simulation using the leapfrog method is numerically stable if and only if the step size is less than twice the square root of the smallest eigenvalue of the covariance matrix \citep{neal2011mcmc}.
Assuming that the local conditional covariance of each mode can be approximated by the inverse of the negative Hessian of the log target density evaluated at the mode, we may tune the step size to be less than the inverse square root of the largest eigenvalue of the Hessian matrix.
In modular ST, the Hessian of the logarithm of the tempered density corresponding to the $k$-th inverse temperature level is given by
\begin{equation}
  \nabla^2 \log \pi_{\beta_k} = (1-\beta_k) \cdot \nabla^2 \log q + \beta_k \cdot \nabla^2 \log \pi.
  \label{eqn:Hessian_convexcomb}
\end{equation}
If the base density $q$ is an isotropic Gaussian with marginal standard deviation $\epsilon_0$, then $-\nabla^2 \log q = \epsilon_0^{-2} I$, suggesting that we use a step size of $\epsilon_0$ for the base density $q$.
The step size $\epsilon$ for the target density $\pi$, corresponding to $\beta_K = 1$, can be chosen to not exceed, approximately, the inverse square roots of the largest eigenvalues of the Hessian matrices across all modes.
In practice, this can be determined by ensuring that the Hamiltonian trajectories are numerically stable for all constrained Markov chains in $\{K\} \times A_i$ at the target temperature level.
Equation~\ref{eqn:Hessian_convexcomb} then suggests that a reasonable step size for the $k$-th inverse temperature level is given by
\begin{equation}
  \epsilon_k := \left\{ (1-\beta_k) \epsilon_0^{-2} + \beta_k \epsilon^{-2} \right\}^{-\frac{1}{2}}.
  \label{eqn:lfsize}
\end{equation}
We used the step sizes given by \eqref{eqn:lfsize} for all numerical examples presented in Section~\ref{sec:numresults}.
The number of numerical integration steps in each MCMC step, which is another tuning parameter in HMC, can be tuned according to the usual guideline of balancing the computational cost per iteration with the average jump distance between successive states.

\subsection{Tuning of temperature levels and mixture weights}\label{sec:tuning}

The mixing speed of simulated tempering depends on the rate of moves between temperature levels, which are in turn determined by the spacing between the inverse temperature levels.
Previous studies on the inverse temperature levels for related tempering methods have suggested tuning them so that the average acceptance probability of temperature swaps is close to a target number between 0 and 1 for parallel tempering, or the ratio of the effective sample size to the ensemble size for each temperature move is close to to another target value between 0 and 1 for tempered sequential Monte Carlo \citep{miasojedow2013adaptive, kone2005selection, atchade2011towards, buchholz2021adaptive}.
We propose a method for tuning the inverse temperature sequence and the mixture weights in modular ST in a similar vein by ensuring that the MH ratios for temperature up or down moves are close to a target value.
We note that, as the relative probabilities of the modes are computed analytically using the eigenvalue method instead of computing the proportion of Monte Carlo draws from the modes, our modular ST is less sensitive to tuning.
However, due to the finite machine precision in the QR factorization, tuning the inverse temperature levels and the mixture weights $\{w_{k,i}\}$ is still important in order to ensure that the estimated stochastic matrix has only one eigenvalue numerically indistinguishable from unity and that the invariant distribution is reliably estimated.

Our method for tuning $\{\beta_k\}$ and $\{w_{k,i}\}$ are as follows.
First, we let $\beta_0 = 0$ and choose $\beta_1$ such that
\begin{equation}
  \beta_1 \cdot \log \left( \frac{\pi(x_0)}{q(x_0)} \right) \approx -1,
  \label{eqn:beta1}
\end{equation}
where $x_0$ is a reference point for the base distribution $q$; for example, if $q$ is the normal distribution, $x_0$ can be the mean of the distribution.
We require that $x_0$ is not very close to any of the mode locations.
The rationale for \eqref{eqn:beta1} is that the proposed temperature moves from $\beta_0 = 0$ and $\beta_1$ should be accepted with reasonable probability.

The intermediate levels between $\beta_1$ and the highest level $\beta_K = 1$ are tuned iteratively using pilot runs.
Let $(\beta_0, \beta_1, \dots, \beta_K)$ be the current sequence of inverse temperature levels.
Construct parallel Markov chains constrained to $A_i \times \{k\}$ for $i \in \{1,\dots, L\}$ and $k \in \{0, \dots, K\}$ of length $n_\text{pilot}$, and evaluate the MH ratio 
\[
  \left( \frac{\pi(X_{k,i}^t)}{q(X_{k,i}^t)} \right)^{\beta_{k'} - \beta_k}
\]
at every iteration where a temperature move is attempted, where $k' = \max(k-1, 0)$ or $\min(k+1, K)$.
When running the pilot runs, all weights $w_{k,i}$ are temporarily set to 1.
For each adjacent pair $(k, k+1)$ for $k=1, \dots ,K-1$ and for each $i \in \{1, \dots, L\}$, denote by $m_{k,\text{up},i}$ the sample median of the logarithm of the MH ratios for temperature moves from $\beta_k$ to $\beta_{k+1}$ on $A_i$, while excluding the first $n_\text{burn-in}$ steps to allow time for each constrained chain to reach stationarity.
Likewise, let $m_{k+1, \text{down}, i}$ denote the sample median of log MH ratios for moves from $\beta_{k+1}$ to $\beta_k$, excluding the burn-in.
If both $m_{k,\text{up},i}$ and $m_{k+1, \text{down},i}$ are small, this suggests that the gap between $\beta_k$ and $\beta_{k+1}$ is too large to enable frequent moves between the two levels on $A_i$.
If this is the case, determine the number of levels to insert between $\beta_k$ and $\beta_{k+1}$ by
\[
  n^\text{insert}_k = \text{Truncate}\left(\max_{i=1,\dots, L} \left[\frac{m_{k,\text{up},i} + m_{k+1, \text{down}, i}}{\log(0.2)} \right], 0, 5\right),
\]
where $[a]$ denotes the largest integer less than or equal to $a$ and we capped the number of inserted temperature levels in each round to 5 for algorithmic stability.
This number is determined such that, once $\{\beta_k\}$ and $\{w_{k,i}\}$ have been tuned, the up or down moves in temperature for every constrained chain in the main MCMC run have average acceptance rates of approximately $\sqrt{0.2}\approx 0.45$.
We insert $n_k^\text{insert}$ intermediate values between $\beta_k$ and $\beta_{k+1}$ with uniform geometric spacing, such that the $j$-th inserted level is given by
\[
  \beta_k \cdot \left( \frac{\beta_{k+1}}{\beta_k} \right)^{\frac{j}{n^\text{insert}_k + 1}}
\]
for $j=1, \dots, n_k^\text{insert}$.
This step is repeated for $k\in \{1, \dots, K-1\}$.
The temperature index $k$ are then updated to include the newly inserted levels.
We iterate this procedure until the number of insertions required for all pairs of adjacent temperature levels are equal to zero.

Once the tuning stage for $\{\beta_k\}$ has been completed, the mixture weights $w_{k,i}$ are determined as follows.
Denoting by $m_{k, \text{up}, i}$ and $m_{k+1, \text{down}, i}$ the median log MH ratios for up or down temperature moves in the last round of tuning $\{\beta_k\}$, we let
\begin{equation}
  \frac{w_{k+1, i}}{w_{k,i}} = \exp\left\{ -\frac{1}{2}(m_{k,\text{up},i} - m_{k+1,\text{down},i}) \right\}
  \label{eqn:w_tuning}
\end{equation}
for $k \in \{0, 1, \dots, K-1\}$ and $i \in \{1, \dots, L\}$.
This step ensures that up and down moves in the inverse temperature level occur with approximately equal probabilities.
We set the mixture weights at the zeroth inverse temperature level all equal to zero: $w_{0,i} = 1$ for $i \in \{1, \dots, L\}$, and determine $w_{k,i}$ for $k\geq 1$ according to \eqref{eqn:w_tuning}.

\subsection{Partitioning algorithm}\label{sec:partition}

The state space needs to be partitioned so that the resulting Monte Carlo estimates have low variance. 
In particular, Assumption~\ref{assum:ir_ap} in Section~\ref{sec:clt} requires that each constrained chain be irreducible.
These requirements can be satisfied by selecting the compartments $A_i$ such that the constrained distributions $\pi_i$ are approximately unimodal for all $i$.
In practice, the partitioning may depend on the available information about the target distribution.

The first step is to identify the mode locations.
Starting from a collection of initial points $\mathbf a_1, \dots, \mathbf a_r$ selected from a broad region where the potential modes are expected, local maxima of the log target density can be found by numerically following the gradient flow, using methods such as gradient ascent or Adam \citep{kingma2017adam}.
Let $\mu_1, \dots, \mu_L$ be the identified modes, where two modes that are sufficiently close to each other are regarded as the same mode.
Next, the partition may be defined as
\[
  A_i = \{ x\in \mathcal X: \text{numerically following the gradient flow starting from } x \text{ reaches } \mu_i \}
\]

To avoid the computational burden of performing numerical optimization for each point, the procedure may be amortized as follows.
For each $i\in \{1,\dots, L\}$, let $S_i$ be the subset of the initial points $\{\mathbf a_1, \dots, \mathbf a_r\}$ from which the search algorithm reached $\mu_i$.
Then we define $A_i$ to be the set of all points in $\mathcal X$ whose closest initial point belongs to $S_i$:
\[
  A_i = \left\{ x \in \mathcal X: \min_{\mathbf a_j \in S_i} d(\mathbf a_j, x) \leq \min_{\mathbf a_{j'} \notin S_i} d(\mathbf a_{j'}, x) \right\}.
\]
The distance function $d(\cdot, \cdot)$ could be the usual Euclidean distance, or if a positive definite matrix $\Sigma$ that approximates the covariance structure of the target density is known, it may be taken to be the Mahalanobis distance
\[
  d(\mathbf x, \mathbf y) = \left\{(\mathbf x - \mathbf y)^\top \Sigma^{-1} (\mathbf x - \mathbf y)\right\}^{1/2}.
\]
We note that an alternative partitioning method based on the spectral clustering of past sample points has been proposed by \citet{basse2016parallel}.

\subsection{Comparison to other methods for multimodal sampling}\label{sec:comparison_topidmixing}
Several other methods have been developed to sample from multimodal distributions with possibly heterogeneous mode scales.
\citet{pompe2020framework} proposed a scheme that attempts direct jumps between the high density regions corresponding to different modes.
This approach approximates the density around each mode by an elliptically shaped distribution, with the covariance matrix adaptively learned on the fly.
\citet{tawn2020weight} proposed a simulated tempering algorithm that directly addresses the issue that modes may have widely different probabilities under tempered distributions when the modes have different scales.
They proposed Hessian adjusted tempering, where in each modal region, the unnormalized tempered density is given by
\[
  \pi(x)^\beta \pi(\mu_i)^{1-\beta}
\]
where $\beta$ denotes the inverse temperature level and $\mu_i$ the location of the $i$-th mode.
This approach can be interpreted as assigning temperature-dependent mixture weights $\pi(\mu_i)^{1-\beta}$ to each mode.
In this regard, Hessian adjusted ST is similar to our modular ST in that mode weights are assigned to facilitate transitions across temperature levels.
However, the probability of accepting proposed jumps in these methods can be low when the modes are not well approximated by elliptical distributions, especially in high dimensions.
By contrast, modular ST can be more robust with respect to irregularities in mode shapes because it does not rely on elliptical approximations.

\section{Estimation of standard error}\label{sec:se_est}
In this section, we propose a method for estimating the standard error of the estimate $\hat \pi h$ obtained from modular MCMC.
We first introduce a method for estimating standard errors for general modular MCMC (Algorithm~\ref{alg:mod_mcmc} and \ref{alg:MHcounter}) and then discuss how the strategy is applied when modular ST (Algorithm~\ref{alg:modular_st}) is used to obtain Monte Carlo estimates from multimodal distributions.

For standard MCMC, standard errors of Monte Carlo estimates can be estimated using block means.
If a Harris recurrent, $V$-uniformly ergodic Markov chain $\{X^t; t\geq 1\}$ has $\pi$ as its invariant distribution and $h$ is a function satisfying $h^2 \leq V$, the Markov chain central limit theorem states that 
\[
  \sqrt n\left(\frac{1}{n} \sum_{i=1}^n h(X^t) - \E_\pi h(X)\right) \Rightarrow \N\{0, \, \sigma^2(h)\}
\]
where the asymptotic variance is given by
\[
  \sigma^2(h) = \text{Var}_{X \sim \pi} h(X) + 2 \sum_{k=1}^\infty \text{Cov}_{X^t \sim \pi}\{h(X^t), h(X^{t+k})\} < \infty
\]
\citep[Theorem~17.0.1]{meyn2009markov}.
Supposing the length of the constructed Markov chain $n$ is a multiple of $s$, we form $n/s$ contiguous blocks of size $s$ as follows:
\[
  \{(X^1, \dots, X^s), (X^{s+1}, \dots, X^{2s}), \dots, (X^{n-s+1}, X^n)\}.
\]
The sample variance of the block means,
\[
  S_\text{block}^2 = \frac{1}{(n/s)-1} \sum_{m=1}^{n/s} (\bar h_m - \bar h)^2,
\]
where $\bar h_m := \tfrac{1}{s} \{h(X^{(m-1)s+1}) + \cdots + h(X^{ms})\}$ and $\bar h = \tfrac{1}{n} \{h(X^1) + \cdots + h(X^n)\}$, gives an estimate of $\sigma^2(h) / s$.
The block size $s$ is chosen such that $\bar h_1, \dots, \bar h_{n/s}$ have autocorrelations close to zero.
The standard error of the MCMC estimate $\hat \pi h = \tfrac{1}{n} \sum_{i=1}^n h(X^i)$ can then be estimated given by
\begin{equation}
  \se ( \hat \pi h ) = \sqrt{\frac{s}{n}} \cdot S_\text{block}.
\label{eqn:block_se}
\end{equation}

\begin{algorithm}[t]
  \caption{Estimation of the standard error of $\hat \pi h$}
  \label{alg:se}
  \begin{algorithmic}[1]
    \Statex \textbf{Input:} Same as Algorithm~\ref{alg:mod_mcmc} or \ref{alg:MHcounter}, and Number of bootstrap replicates $B$; Block size $s$
    \Statex \textbf{Output:} Estimated standard error, $\se(\hat \pi h)$
    \Statex \textit{Run Algorithm~\ref{alg:mod_mcmc} or \ref{alg:MHcounter} and then do the following}
    \For {each $i \in 1\cl L$}
      \State Let $\bar h^m_i = \tfrac{1}{s} \{ h(X_i^{(m-1)s+1} + \cdots + h(X_i^{ms}) \}$ be the $m$-th block sum of $h(X_i^t)$
      \State Let $\se(\hat \pi_i h)$ be $\sqrt{s / n_i}$ times the sample standard deviation of $\{\bar h^m_i: m =1, \dots, n_i/s\}$
      \State For each $j \neq i$, let $\Delta_{ij}^m = C_{ij}^{ms} - C_{ij}^{(m-1)s}$ be the $m$-th block transition counter, $m = 1,\dots, n_i/s$
      \State Let $\hat \Sigma_i$ be $\tfrac{1}{n_i s}$ times the sample variance-covariance matrix of the $L-1$ dimensional vectors $\{(\Delta_{ij}^m)_{j\neq i}; m=1, \dots, n_i/s\}$
      \State Draw $(\hat Q_{ij}^b)_{j\neq i}$ from the multivariate normal distribution $N\{(\hat Q_{ij})_{j\neq i}, \hat \Sigma_i\}$, $b \in 1\cl B$
      \State Let $\hat Q_{ii}^b = 1 - \sum_{j\neq i} \hat Q_{ij}^b$, $b \in 1\cl B$
    \EndFor
    \State Let $\hat {\mathbf p}^b = (\hat p_i^b)_{i \in 1:L}$ be the left eigenvector of $(\hat Q_{ij}^b)_{i\in 1:L, j\in 1:L}$ corresponding to the unit eigenvalue for each $b \in 1\cl B$
    \State Let $\widehat{\pi(A_i)}^b = (\hat p_i^b / w_i) / \sum_{j=1}^L (\hat p_j^b / w_j)$ for each $i\in 1\cl L$ and $b \in 1\cl B$
    \State Let $F_1 = \sum_{i=1}^L \left\{ \frac{1}{B} \sum_{b=1}^B (\widehat{\pi(A_i)}^b)^2 \right\} \cdot \se(\hat \pi_i h)^2$ (Equation~\ref{eqn:se_first_term})
    \State Let $F_2$ be the sample variance of $\{\hat \pi^b h := \sum_{i=1}^L \widehat{\pi(A_i)}^b \cdot \hat \pi_i h: b \in 1\cl B\}$
    \State Let $\se(\hat \pi h) = F_1 + F_2$
  \end{algorithmic}
\end{algorithm}

In modular MCMC, the standard error of the Monte Carlo estimate $\hat \pi h$ depends on both the standard errors of the estimated constrained expectations $\hat \pi_i h$ for $i \in \{1,\dots, L\}$ and those of the estimated compartment probabilities $(\widehat{\pi(A_i)})_{i\in 1:L}$.
The standard error of $\hat \pi_i h = \tfrac{1}{n_i} \sum_{t=1}^{n_i} h(X_i^t)$ for each $i \in \{1, \dots, L\}$ can be estimated using block means according to \eqref{eqn:block_se}.
We denote this estimated standard error by $\se ( \hat \pi_i h )$.
To estimate the standard errors of $(\widehat{\pi(A_i)})_i$, we first estimate the joint variance-covariance matrix of the transition probabilities $\hat Q_{ij}$ for $i \neq j$.
Consider block sums of transition counters from compartment $i$ to $j$ of size $s$:
\[
  \Delta_{ij}^1 := C_{ij}^s, \quad \Delta_{ij}^2 := C_{ij}^{2s} - C_{ij}^s, \quad \cdots, \quad \Delta_{ij}^{n_i/s} := C_{ij}^{n_i} - C_{ij}^{n_i-s}.
\]
Then we have $\hat Q_{ij} = \tfrac{1}{n_i} (\Delta_{ij}^1 + \cdots + \Delta_{ij}^{n_i/s})$.
Assuming that $\Delta_{ij}^1, \dots, \Delta_{ij}^{n_i/s}$ are approximately independent, we can estimate the variance of $\hat Q_{ij}$ by
\[
  \widehat{\text{Var}} (\hat Q_{ij}) = \frac{1}{n_i s} \widehat{\text{Var}}(\Delta_{ij}),
\]
where $\widehat{\text{Var}}(\Delta_{ij})$ denotes the sample variance of $\Delta_{ij}^1, \dots, \Delta_{ij}^{n_i/s}$.
Similarly, the covariance between $\hat Q_{ij}$ and $\hat Q_{ij'}$, where $i\neq j$ and $i\neq j'$, can be estimated by
\[
  \widehat{\text{Cov}}(\hat Q_{ij}, \hat Q_{ij'}) = \frac{1}{n_i s} \widehat{\text{Cov}}(\Delta_{ij}, \Delta_{ij'}),
\]
where $\widehat{\text{Cov}}(\Delta_{ij}, \Delta_{ij'})$ denotes the sample covariance of $\Delta_{ij}^m$ and $\Delta_{ij'}^m$ across $m=1, \dots, n_i/s$.

We propose a bootstrap method to simulate the variability in the estimated compartment probabilities as follows.
We repeat the following procedure for each $i\in \{1,\dots,L\}$, as the constrained chains are constructed independently.
We generate a bootstrap sample $\{(\hat Q_{ij}^b)_{j\neq i} : b \in 1 \cl B\}$ from the multivariate normal distribution with mean $(\hat Q_{ij})_{j \neq i}$ and variance-covariance matrix with diagonal entries $\widehat{\text{Var}}(\Delta_{ij})$ and off-diagonal entries $\widehat{\text{Cov}}(\Delta_{ij}, \Delta_{ij'})$.
Each entry $\hat Q_{ij}^b$ for $j \neq i$ is then truncated to be nonnegative by updating
\[
  \hat Q_{ij}^b \gets \max(\hat Q_{ij}^b , 0),
\]
and then if $\sum_{j\neq i} \hat Q_{ij}^b$ is greater than 1, update each $Q_{ij}^b$ by
\[
  \hat Q_{ij}^b \gets \frac{\hat Q_{ij}^b}{\sum_{j\neq i} \hat Q_{ij}^b}.
\]
Lastly, the diagonal entries are computed as $\hat Q_{ii}^b = 1 - \sum_{j \neq i} \hat Q_{ij}^b$.

For each $b$, we generate a matrix $\hat Q^b = (\hat Q_{ij}^b)_{i \in 1:L, j\in 1:L}$ by combining $\hat Q_{ij}^b$ across $i \in \{1,\dots, L\}$.
Then, the left eigenvector $\hat {\mathbf p}^b = (\hat p_i^b)_{i \in 1:L}$ of $\hat Q^b$ corresponding to the unit eigenvalue is found using the QR decomposition, as introduced in Section~\ref{sec:modmcmc}.
The bootstrap samples for the estimated compartment probabilities are computed as
\[
  \widehat{\pi(A_i)}^b = \frac{ \hat p_i^b / w_i }{ \sum_{j=1}^L (\hat p_j^b / w_j) }, \qquad b \in 1\cl B.
\]

The variance of the modular MCMC estimate $\hat \pi h$ (Equation~\ref{eqn:pihat}) can be approximated using the estimated standard errors of $(\hat \pi_i h)_i$ and $(\widehat{\pi(A_i)})_i$ and the following identity:
\begin{equation}
  \text{Var}( \hat \pi h )
  = \E\left[ \text{Var}\left\{\hat \pi h \middle| (\widehat{\pi(A_i)})_{i\in 1:L}\right\} \right]
  + \text{Var}\left[ \E\left\{\hat \pi h \middle| (\widehat{\pi(A_i)})_{i\in 1:L}\right\} \right].
  \label{eqn:var_blackwell}
\end{equation}
The conditional variance in the first term on the right hand side of \eqref{eqn:var_blackwell} can be approximated as
\[
  \text{Var}\left\{\hat \pi h \middle| (\widehat{\pi(A_i)})_{i\in 1:L}\right\}
  = \text{Var}\left\{ \sum_{i=1}^L \widehat{\pi(A_i)} \cdot \hat \pi_i h \middle| (\widehat{\pi(A_i)})_{i=1}^L \right\}
  \approx \sum_{i=1}^L \widehat{\pi(A_i)}^2 \cdot \se(\hat \pi_i h)^2,
\]
since the constrained chains for $i \in \{1,\dots, L\}$ are constructed independently.
Thus we approximate the first term on the right hand side of \eqref{eqn:var_blackwell} by
\begin{equation}
  \E\left[ \text{Var}\left\{ \hat \pi h \middle| (\widehat{\pi(A_i)})_{i\in 1:L} \right\} \right]
  \approx \sum_{i=1}^L \left\{ \frac{1}{B} \sum_{b=1}^B \left(\widehat{\pi(A_i)}^b \right)^2 \right\} \cdot \se(\hat \pi_i h)^2.
  \label{eqn:se_first_term}
\end{equation}
The second term on the right hand side of \eqref{eqn:var_blackwell} is approximated by the sample variance of the $B$ bootstrap estimates given by
\[
  \hat \pi^b h := \sum_{i=1}^L \widehat{\pi(A_i)}^b \cdot \hat \pi_i h, \qquad b \in 1\cl B.
\]
Algorithm~\ref{alg:se} summarizes the steps for estimating the standard error of modular MCMC estimates.

The same method can be applied to estimate the standard errors of Monte Carlo estimates obtained from modular simulated tempering (Algorithm~\ref{alg:modular_st}).
The estimates of the constrained expectations are given by
\[
  \hat \pi_i h = \frac{1}{n_{K,i}} \sum_{t=1}^{n_{K,i}} h(X_{K,i}^t) \qquad \text{for } i \in 1\cl L.
\]
For modular ST, we have $\hat Q_{(k,i) \to (k',j)}$ equal to 0, unless (a) $k = k'$ or (b) $i=j$ and $|k-k'|=1$.
Thus, for each $k \in \{0, \dots, K\}$ and $i \in \{1,\dots, L\}$, bootstrap samples $(\hat Q^b_{(k,i)\to (k',j)})_{k',j}$ are generated from the multivariate normal distribution of dimension $L+1$ or $L+2$, depending on whether $k \in \{0,K\}$.
The bootstrap estimates of the compartment probabilities are calculated using the entries of the eigenvector of each $\hat Q^b$ corresponding to the $K$-th inverse temperature level,
\[
  \widehat{\pi(A_i)}^b = \frac{\hat p_{K,i}^b / w_{K,i}}{\sum_{j=1}^L \hat p_{K,j}^b / w_{K,j}},
\]
where $\hat{\mathbf p}^b = (\hat p_{k,i}^b)_{k\in 0:K, i \in 1:L}$ satisfies $(\hat{\mathbf p}^b)^\top \hat Q^b = (\hat{\mathbf p}^b)^\top$.
The standard error of $\hat \pi h$ is estimated using Equation~\ref{eqn:var_blackwell}: its square is given by the sum of
\[
  \sum_{i=1}^L \left\{ \frac{1}{B} \sum_{b=1}^B \left( \widehat{\pi(A_i)}^b \right)^2 \right\} \cdot \se(\hat \pi_i h)^2
\]
and the sample variance of
\[
  \hat \pi^b h = \sum_{i=1}^L \widehat{\pi(A_i)}^b \cdot \hat \pi_i h \qquad \text{for } b \in 1\cl B.
\]

\section{Numerical results}\label{sec:numresults}
In this section, we apply modular ST to obtain Monte Carlo estimates of expectations with respect to strongly multimodal target distributions.
As a first example, we use mixtures of Gaussian densities with varying degrees of differences in the widths of the mixture components.
This example highlights the capability of our modular approach to accurately estimate the relative probabilities of modes of different sizes---an issue that poses significant challenges for tempering-based sampling methods.
We compare our method with parallel tempering (PT) and tempered sequential Monte Carlo (TSMC).

In the second example, we demonstrate that modular ST can successfully sample from the Bayesian posterior distribution arising from a regression model that encourages sparsity through a spike-and-slab prior.
This example shows that our approach can tackle challenging computational problems of practical importance in statistical data analysis.

All code used to generate the numerical result is available at \url{https://github.com/joonhap/modularMCMC_code}.

\subsection{Mixture of Gaussian components of different scales}\label{sec:mixGaussian}
To demonstrate modular ST (Algorithm~\ref{alg:modular_st}) when the target distribution consists of modes of different scales, we considered $d$-dimensional Gaussian mixture distributions given by
\[
  \frac{1}{2} \mathcal N(\mathbf \mu_1, \sigma_1^2 I_d) + \frac{1}{2} \mathcal N(\mathbf \mu_2, \sigma_2^2 I_d).
\]
Each entry of the two mean vectors, $\mathbf \mu_1$ and $\mathbf \mu_2$, was drawn independently and uniformly from $[-10, 10]$:
\[
  \mathbf \mu_1, \mathbf \mu_2 \overset{iid}\sim \text{Uniform}[-10, 10]^{\otimes d}
\]
The two mixture components had isotropic covariance matrices with marginal standard deviations given by
\[
  \sigma_1 = 0.1, \quad \sigma_2 = \rho^{\frac{1}{d}} \sigma_1,
\]
where $\rho$ denotes the scale ratio, which was varied over $\{1, 10, 100, 1000\}$.
Thus, the volume of the high-density region of the second mixture component was approximately $\rho$ times greater than that of the first mixture component.

We implemented three methods---modular ST, adaptive PT, and adaptive TSMC.
The inverse temperature levels $\{\beta_k\}$ for modular ST (Algorithm~\ref{alg:modular_st}) were tuned according to the strategy introduced in Section~\ref{sec:tuning}.
The base distribution $q$ for simulated tempering was chosen to be the normal distribution with mean $\mathbf 0 \in \mathbb R^d$ and covariance $20^2 \cdot I_d$.
For this Gaussian mixture example, the method of partitioning the state space based on numerically following gradient flows, as introduced in Section~\ref{sec:partition}, can be approximated by a simpler procedure that defines the partition by comparing the densities of the two mixture components as follows:
\[
  A_i = \{ x: \phi(x; \mathbf \mu_i, \sigma_i^2 I_d) > \phi(x; \mathbf \mu_j, \sigma_j^2 I_d) \text{ for } j\neq i\}, \quad i=1, 2,
\]
where $\phi(x; \mathbf \mu_i, \sigma_i^2 I_d)$ denotes the density of each multivariate Gaussian mixture component.
We used an HMC kernel to construct each Markov chain constrained to $\{k\} \times A_i$, using a leapfrog step size determined by \eqref{eqn:lfsize} with the target-level step size $\epsilon = 0.1$ and the base-level step size $\epsilon_0 = 10$.
Each iteration of HMC consisted of ten leapfrog steps.
All parallel Markov chains were constructed to have length $20,000$.

Our implementation of parallel tempering targeted the joint distribution of independent tempered distributions having densities proportional to $\pi^{1/\alpha_k}$ where $1=\alpha_0 < \alpha_1 < \cdots < \alpha_K$ represent different temperature levels.
For exchanging states between adjacent temperature levels, we adopted the deterministic even-odd exchange scheme proposed by \citet{syed2022nonreversible}.
The temperature levels $\{\alpha_k\}$ were tuned adaptively, based on the method proposed by \citet{miasojedow2013adaptive}, with a target swap rate of 0.234, following the suggestion of \citet{kone2005selection} and \citet{atchade2011towards}.
To ensure that transitions between modes could occur frequently at the top temperature level, we adaptively added new chains until the top-level chain was able to sample points outside $[-20, 20]^d$ with reasonable frequency.
At each temperature level, an HMC kernel consisting of 50 leapfrog steps was applied in each iteration, where the leapfrog step size was tuned adaptively.
Each parallel chain had length $20,000$.
Further implementation details can be found in the supplementary material.

Tempered sequential Monte Carlo (TSMC) is an algorithm that sequentially transforms an ensemble of particles to represent a sequence of intermediate distributions that bridge a base distribution and the target distribution \citep{neal2001annealed}.
We employed intermediate distributions with density proportional to $\pi(x)^{\beta_k} q(x)^{1-\beta_k}$, like the ones used in modular ST.
As TSMC moves from $\beta_0 = 0$ to $\beta_K = 1$, the ensemble of particles is iteratively resampled according to importance weights to reflect the changes in the tempered distributions and perturbed to replenish particle diversity.
At each perturbation step, we applied an HMC kernel with the corresponding tempered distribution as its invariant distribution five times in succession, where each application consisted of ten leapfrog steps with adaptively tuned step sizes.
The inverse temperature levels $\{\beta_k\}$ were tuned such that the effective sample size at each resampling step was approximately half the ensemble size.
Upon reaching the last inverse temperature level $\beta_K = 1$, the algorithm produces a weighted sample $\{(w_j, x_j): j\in 1\cl J\}$ representing the target distribution $\pi$.
We used $J=20,000$ particles.

We obtained Monte Carlo samples using the three methods for $d \in \{1, 5, 10, 100\}$ and $\rho \in \{1, 10, 100, 1000\}$, resulting in a total of sixteen different settings.
For each method, the experiment was replicated 40 times under each setting.
In order to compare the bias and variance of the resulting Monte Carlo estimates, we considered a test function
\[
  h(x) = \mathbf 1[\Vert x - \mathbf \mu_1\Vert < \Vert x - \mathbf \mu_2\Vert ].
\]
Except in the two cases $(d,\rho) = (1,100)$ and $(1, 1000)$, the two mixture components have well-separated densities, and thus the expectation $\pi h$ is very close to $\tfrac{1}{2}$, the mixture probability of the first component.
Since sampling of each Gaussian mixture component is straightforward in this example, the Monte Carlo variances of various estimates are essentially determined by the Monte Carlo variance of the estimates of the mixture weights.

The computation time was comparable for all three methods.
The total numbers of leapfrog steps performed, regarded as an approximate indicator of the computational cost, are provided in the supplementary material.
Due to the fact that parallel tempering requires frequent communication between parallel chains to exchange sample points, PT has limitations in using multiple cores efficiently.
In our experiments, PT took about twice as long to run as the other two methods, even though it performed fewer leapfrog steps overall.
Both modular ST and TSMC are well suited to take advantage of the available parallelism across multiple computing units.

\begin{figure}[tp]
  \centering
  \includegraphics[width=6.5in]{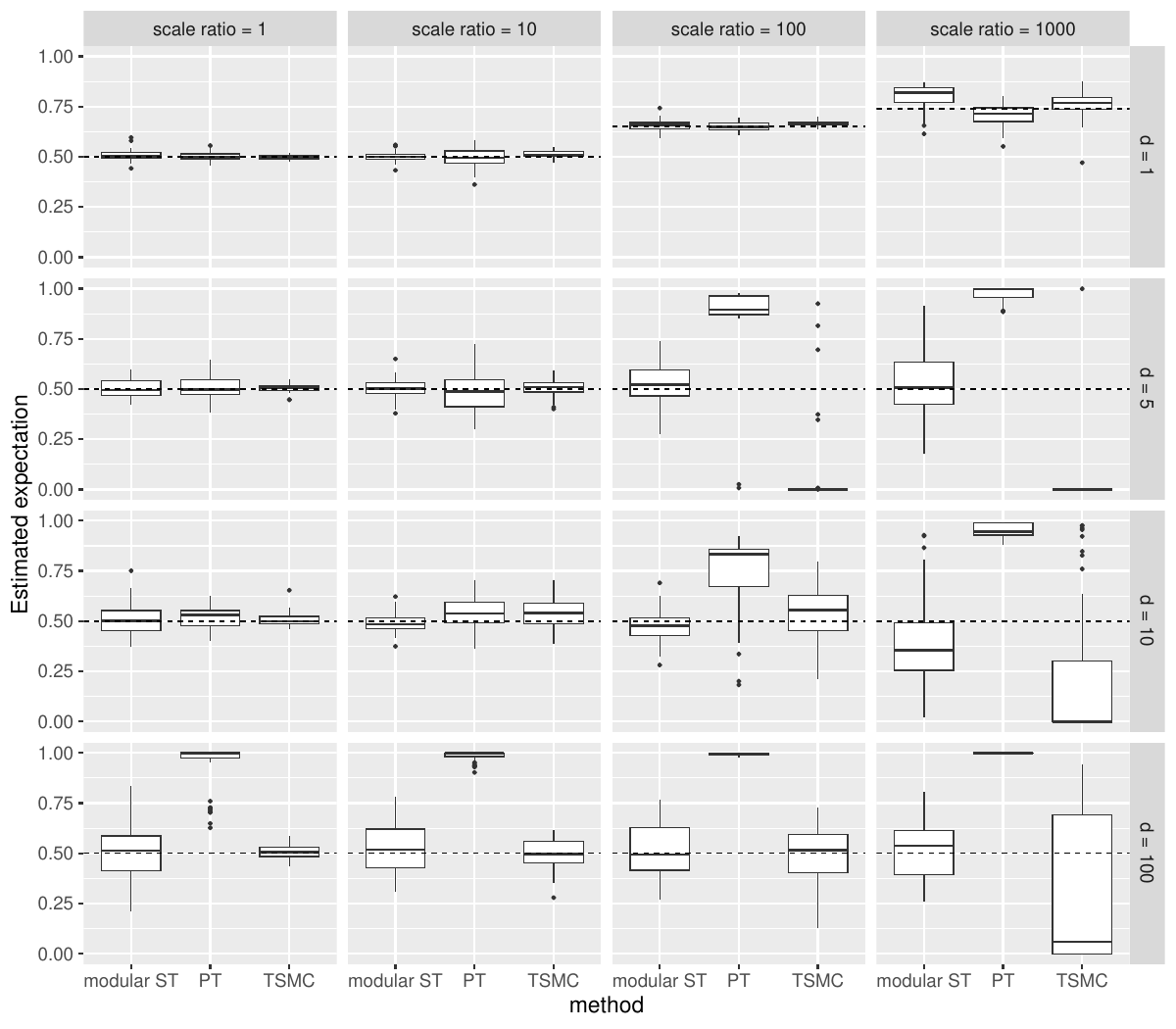}
  \caption{Boxplots of estimates of $\pi h$---the probability that a random draw from the target distribution is closer to the first mode than to the second mode---for three methods under varying dimension ($d$) and scale ratio between modes ($\rho$).
    Each boxplot shows the distribution of estimates across 40 replications.
    The horizontal dashed lines indicate the theoretically expected values.}
  \label{fig:comparison}
\end{figure}

Figure~\ref{fig:comparison} shows the estimates of the expectation $\pi h$ produced by the three methods under different settings.
Modular ST exhibits low or moderate bias and variance in all combinations of $d$ and $\rho$.
In contrast, both PT and TSMC show increasing bias as $\rho$ increases from 1 to 1000, and this behavior is more pronounced in higher dimensions.
Especially when $\rho = 1000$, both PT and TSMC produce extremely inaccurate estimates in dimensions $d \geq 5$.
For relatively large $\rho$ and $d$, TSMC tends to sample substantially more particles from the second mode, which has much greater probability than the first mode at high temperature levels.
The low proportion of particles near the first mode at higher temperature levels does not increase to the theoretically expected proportion of $\tfrac{1}{2}$ at the lowest temperature level as the algorithm transforms the particle ensemble through intermediate resampling steps.

In parallel tempering, all chains were initialized at points near the first mode.
When the scale difference between the modes was large, the samples near the second mode at high temperatures failed to trickle down to the bottom temperature level in dimensions $d \geq 5$.
Notably, transition between modes at the lowest temperature level occurred very infrequently even when the two modes had the same scale (i.e., $\rho=1$) in dimension $d=100$.
This illustrates the unfavorable dimension scaling of parallel tempering even when it is adaptively tuned.
Compared to the other methods, however, modular ST exhibits moderately favorable dimension scaling and robust sampling accuracy under heterogeneous mode scales.

\begin{figure}[tp]
  \centering
  \includegraphics[width=6in]{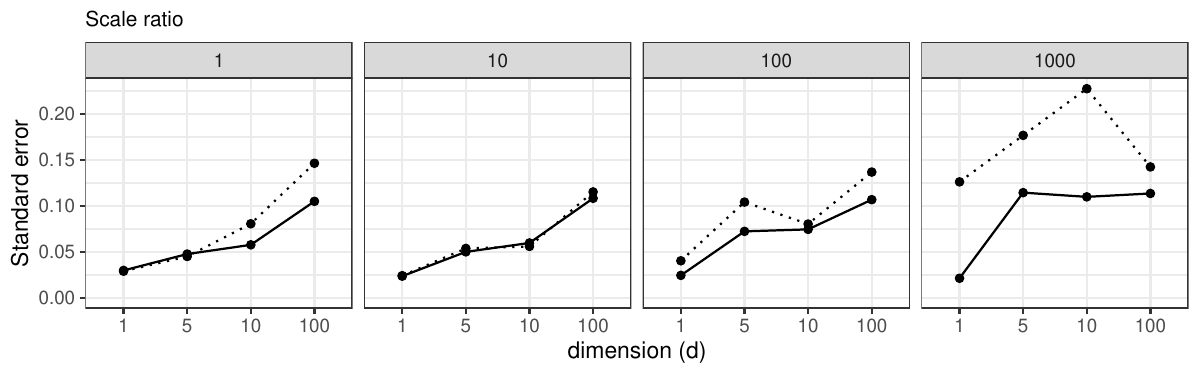}
  \caption{Estimated standard error (solid) of $\widehat{\pi}(A_1)$ by Algorithm~\ref{alg:se} for modular ST for the mixture of Gaussian example under varying dimension ($d$) and scale ratio between modes ($\rho$).
    Dotted lines indicate the sample standard deviation of estimates of $\widehat{\pi}(A_i)$ across 40 replications.
  }
  \label{fig:se_est}
\end{figure}

For modular ST, we estimated the standard error of the estimated probability of the first mode, $\se\{\widehat{\pi}(A_1)\}$, using Algorithm~\ref{alg:se}.
Figure~\ref{fig:se_est} shows the standard error for varying $d$ and $\rho$, along with the sample standard deviation of the estimates across forty replications.
For $\rho = 1$, 10, and 100, the estimated standard errors were reasonably close to the empirical Monte Carlo variability of the estimates.
Even when the two modes had greatly different scales ($\rho = 1000$), the standard errors estimated by Algorithm~\ref{alg:se} had the same order of magnitude as the empirical standard deviations of the estimates.

\subsection{Sparse linear regression with a spike-and-slab prior}\label{sec:sparse_reg}
We considered the following Bayesian sparse linear regression problem for $d=30$ covariates and a response variable.
The model is defined as
\begin{equation}
  Y_i = \sum_{j=1}^d b_j x_{i,j} + \epsilon_i, \qquad \epsilon_i \overset{iid}\sim \mathcal N(0,1), \qquad i=1,\dots, 100.
  \label{eqn:linmodel}
\end{equation}
To induce sparsity in the estimated coefficient vector $b = (b_1, \dots, b_d)^\top$, we imposed a spike-and-slab prior, where each component is independently assigned a mixture distribution
\[
  b_j \sim 10^{-4} \mathcal N(0, \sigma_1^2) + (1-10^{-4}) \mathcal N(0, \sigma_2^2)
\]
with $\sigma_1 = 10$ and $\sigma_2 = 1$.
The log posterior density for $b$ was then given by
\[
  \log \pi(b) = -\frac{1}{2} \Vert Y - Xb \Vert^2 + \log \left( \frac{10^{-4}}{\sigma_1^d} e^{-\frac{1}{2\sigma_1^2} \Vert b\Vert^2} + \frac{1-10^{-4}}{\sigma_2^d} e^{-\frac{1}{2\sigma_2^2} \Vert b\Vert^2} \right) + const.
\]
where $Y = (y_1, \dots, y_{100})^\top$ and $X \in \mathbb R^{100\times 30}$ denotes the design matrix with entries $x_{i,j}$.
Each $x_{i,j}$ for $i\in 1\cl 100$ and $j\in \{1, 3, 4, \dots, 30\}$ was randomly drawn from the standard normal distribution, while we set $x_{\cdot,1} = x_{\cdot,2}$ so that the posterior distribution is strongly multimodal.

The data was generated using the model \eqref{eqn:linmodel}, where the first four coefficients were
\[
  (b_1, b_2, b_3, b_4) = (10, 15, -10, -15)
\]
and all other parameters were equal to zero.
Due to the collinearity $x_{\cdot,1} = x_{\cdot,2}$, any $b$ having the same value of $b_1 + b_2$ gives the same likelihood.
However, the spike-and-slab prior puts more probability mass where only one of $b_1$ and $b_2$ has large absolute value.
As a result, the posterior has two well separated modes where the first two entries are near $(b_1, b_2) = (25, 0)$ and $(0, 25)$.

We applied modular ST (Algorithm~\ref{alg:modular_st}) to sample from the posterior distribution.
Two modes, denoted by $\hat b^{(1)}$ and $\hat b^{(2)}$, were discovered using gradient ascent.
The parameter space was partitioned into two sets,
\[
  A_1 = \left\{ b \in \mathbb R^{30} : \Vert b - \hat b^{(1)} \Vert < \Vert b - \hat b^{(2)} \Vert \right\}, \quad \text{and} \quad A_2 = \mathbb R^{30} \setminus A_1.
\]

We used an HMC kernel to construct each constrained Markov chain.
In order to increase the efficiency of HMC, we approximated the negative Hessian of the log posterior density by
\[
  -\nabla^2 \log \pi(b) \approx \Omega := X^\top X + \frac{1}{\sigma_1^2} I_d
\]
and reparametrized by
\[
  c := \Omega^{1/2} b
\]
where $\Omega^{1/2}$ denotes the square-root matrix of the symmetric matrix $\Omega$.
Every step of HMC consisted of ten leapfrog steps with the step size determined by \eqref{eqn:lfsize} with $\epsilon_0 = 10$ and $\epsilon = 0.5$.
The base distribution $q$ was the normal distribution with mean $0 \in \mathbb R^d$ and covariance $20^2 I_d$.
The sequence of inverse temperature levels $\{\beta_k\}$ were tuned using iterative pilot runs as introduced in Section~\ref{sec:tuning}. 
We carried out 20 replicated numerical experiments.
Across these replications, the mean number of temperature levels was 139.6 and the standard deviation was 10.5.
Each of the parallel chains was constructed to have length 20000.

\begin{table}
  \centering
  \begin{tabular}{lllllll}\toprule
    & $b_1$ & $b_2$ & $b_3$ & $b_4$ & $b_5$ & $b_6$ \\\midrule
    mode~1 & 24.63 (0.03) & 0.28 (0.03) & -9.99 (0.00) & -14.91 (0.00) & -0.11 (0.00) & -0.04 (0.00) \\
    mode~2 & 0.33 (0.07) & 24.59 (0.07) & -9.99 (0.00) & -14.91 (0.00) & -0.10 (0.00) & -0.04 (0.00) \\\bottomrule
  \end{tabular}
  \caption{
    The estimated posterior means of the first six coefficients from the two constrained Markov chains at the bottom temperature level ($\beta_K = 1$), averaged over twenty replications.
    The standard deviations are indicated between the parentheses.
  }
  \label{tab:lin_coef}
\end{table}

Table~\ref{tab:lin_coef} gives the estimated posterior mean of the first six entries of the parameter vector from each of the two constrained Markov chains at the target temperature level $\beta_K = 1$.
As expected, only one of the posterior means of $b_1$ and $b_2$ was significantly different from zero, and its value was approximately equal to the sum of the true coefficients (namely, $b_1 + b_2 = 25$).
The posterior means of all other parameters were close to the true coefficients as well.

\begin{table}
  \centering
  \begin{tabular}{llll}
    \toprule
    $\widehat{\pi(A_1)}$ & mean & standard deviation & estimated standard error \\\midrule
    all                  & 0.51 & 0.20 & 0.17 \\
    excluding one replication & 0.47 & 0.07 & 0.03\\\bottomrule
  \end{tabular}
  \caption{
    The mean, standard deviation, and estimated standard error of the estimated normalized weight of the first mode, $\widehat{\pi(A_1)}$, across 20 replications.
    The first row reports the values across all 20 replications, and the second reports the values excluding the one case in which $\widehat{\pi(A_1)}$ was greater than 1.
  }
  \label{tab:sparse_prob}
\end{table}

Table~\ref{tab:sparse_prob} shows the mean and standard deviation of the estimated weight of the first mode, $\widehat{\pi(A_1)}$, across 20 replications, as well as its standard error estimated using Algorithm~\ref{alg:se}.
In one of the 20 replications, the left eigenvector of the estimated stochastic matrix $\hat Q$ corresponding to the unit eigenvalue had a negative entry, resulting in $\widehat{\pi(A_1)} > 1$ and $\widehat{\pi(A_2)} < 0$.
This numerical anomaly likely occurred due to outliers in the matrix $\hat Q$.
Excluding this single case, the mean of $\widehat{\pi(A_1)}$ was 0.47, with a standard deviation of 0.07.
Due to the symmetry between the two posterior modes in this example, the exact probability is $\pi(A_1) = 0.5$.
The Monte Carlo estimates of the weight of the first mode was reasonably close to this exact probability.
This was somewhat surprising given that there were approximately 140 tuned inverse temperature levels and more than 100 log units of difference among the tuned mixture weights $w_{k,i}$.
Moreover, the estimated standard error of $\widehat{\pi(A_1)}$ was on the same order of magnitude as the empirical standard deviation of the estimates.
These results suggest that modular ST can reliably estimate posterior expectations and quantify the associated uncertainties in computationally challenging Bayesian inference problems.

\section{Conclusion}

We developed a modular approach to MCMC, in which parallel Markov chains are constructed on non-overlapping regions of the state space.
A key step is the estimation of the probability of each region, which is carried out through the eigendecomposition of a stochastic matrix whose entries indicate the transition rates between regions under a global Markov kernel.
In addition to the computational advantage of being able to use multiple computing units in parallel, modular MCMC can reduce the variance of Monte Carlo estimates.
We focused on the estimation of expectations with respect to multimodal target distributions by applying the modular approach to a state space augmented with an inverse temperature variable.
This modular simulated tempering algorithm enables the estimation of probabilities of modes having different scales, thereby addressing an important challenge for tempering-based sampling methods.

Another potential use case of modular MCMC is the estimation of small probabilities, since the probabilities of partition components are estimated by solving an algebraic equation rather than computing the empirical proportion of sample draws in the region of interest.

A likely future research direction is improving the dimension scaling of modular ST.
A tempered HMC algorithm that scales to high dimensions has been developed \citep{park2024sampling}, but it shares with other tempering methods the challenge of slow mixing when the modes have different scales.
A method that enables efficient sampling from a broad range of multimodal distributions in high dimensional settings would be useful for many applications in statistics, computational physics, and other areas.

\section{Acknowledgments}
The author gratefully acknowledges support from the General Research Fund at the University of Kansas and from the Don and Pat Morrison Foundation.

\section{Data Availability Statement}\label{data-availability-statement}

The source code used to generate the numerical results has been made available at \url{https://github.com/joonhap/modularMCMC_code}.

\includepdf[pages=-]{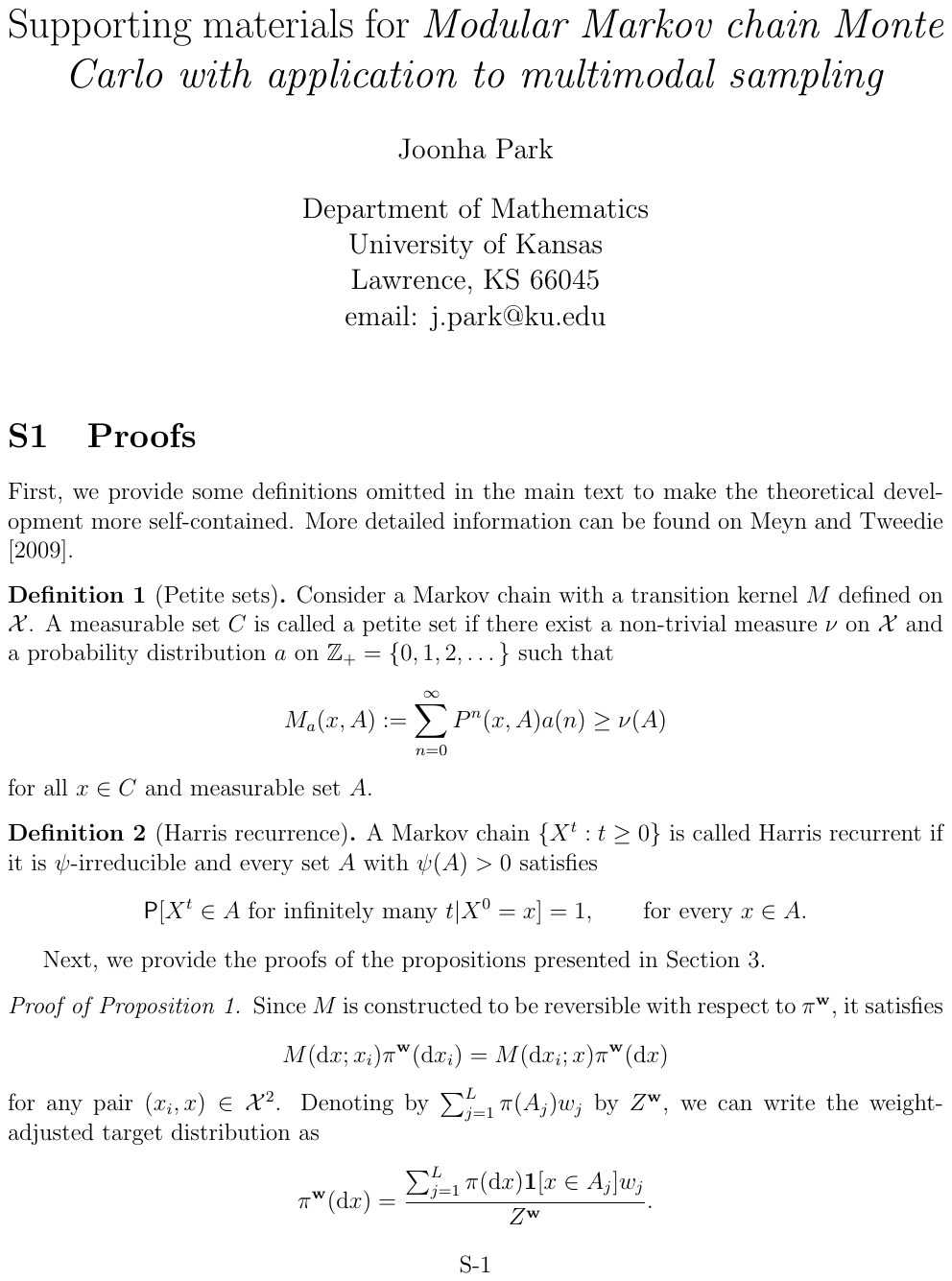}

\end{document}